\documentclass[sigconf,authorversion]{acmart}
\usepackage{paralist}
\usepackage{acronym}
\usepackage{enumitem}
\usepackage{subcaption}
\usepackage{dirtytalk}
\usepackage{amsmath}
\usepackage[group-separator={,},output-decimal-marker={.}]{siunitx}
\usepackage{usebib}
\bibinput{references}
\usepackage{todonotes}
\usepackage{csquotes}
\usepackage{svg-extract}
\usepackage[capitalise,noabbrev,nameinlink]{cleveref}
\usepackage{tabularx}
\usepackage{threeparttable}
\usepackage{arydshln}
\usepackage{array}
\usepackage{placeins}
\usepackage{overpic}
\usepackage{transparent}
\usepackage{xfp}
\usepackage{url}
\usepackage{balance}

\newcommand{\shortlink}[1]{\href{https://#1}{\path{#1}}}

\makeatletter
\AtBeginDocument
 {
   \def\ltx@label#1{\cref@label{#1}}
   \def\label@in@display@noarg#1{\cref@old@label@in@display{#1}}
\def\label@in@mmeasure@noarg#1{%
    \begingroup%
      \measuring@false%
      \cref@old@label@in@display{#1}
    \endgroup}%
 } %

\newcommand{\as}[2][]{%
    \def\@tempa{#1}%
    \def\@tempb{link=true}%
    \ifx\@tempa\@empty
        \texttt{AS#2}%
    \else\ifx\@tempa\@tempb
        \href{https://ipinfo.io/AS#2}{\texttt{AS#2}}%
    \else
        \texttt{AS#2}%
    \fi\fi
}
\makeatother

\newcommand{\boxlink}[3]{%
    \edef\height{\fpeval{414.4}}
    \edef\width{\fpeval{351.931938}}
    \edef\x{#1}
    \edef\y{#2}

    \edef\x{\fpeval{(\x-3) / \height * 100}}
    \edef\y{\fpeval{(\height - (\y+3)) / \height * 100}}

    \put(\x,\y){%
        \href{#3}{%
            \transparent{0}\color{red}\rule{4pt}{4pt}%
        }%
    }
}

\newcommand{\itwquote}[1]{%
    \say{\textit{#1}}%
}

\setlist[itemize]{leftmargin=*}
\setlist[enumerate]{leftmargin=*}

\newlist{rqenumerate}{enumerate}{2}
\setlist[rqenumerate,1]{%
    label=\bfseries\itshape RQ~\arabic*,
    ref=\arabic*,
    labelindent=0pt, itemsep=0.5em}

\crefname{rqenumeratei}{RQ}{RQs}

\newlist{contribenum}{enumerate}{2}
\setlist[contribenum,1]{%
    label=\bfseries\itshape C~\arabic*,
    ref=\arabic*,
    labelindent=0pt, itemsep=0.5em}

\crefname{contribenumi}{C}{Cs}

\copyrightyear{2026}
\acmYear{2026}
\setcopyright{cc}
\setcctype{by}
\acmConference[WPES '26]{25th Workshop on Privacy in the Electronic Society}{November 15--19, 2026}{The Hague, Netherlands}
\acmBooktitle{25th Workshop on Privacy in the Electronic Society (WPES '26), November 15--19, 2026, The Hague, Netherlands}
\acmDOI{10.1145/3847192.3847370}
\acmISBN{979-8-4007-3026-9/2026/11}

\begin{document}

\title{Sociotechnical Aspects of Tor Relay Rejection}


\author{Jules Dejaeghere}
\authornote{Shared first author.}
\orcid{0000-0002-4970-3730}
\affiliation{%
  \institution{University of Namur}
  \city{Namur}
  \country{Belgium}}
\email{jules.dejaeghere@unamur.be}

\author{Lionel Goffaux}
\authornotemark[1]
\orcid{0009-0004-1331-7215}
\affiliation{%
  \institution{University of Namur}
  \city{Namur}
  \country{Belgium}}
\email{lionel.goffaux@unamur.be}

\author{Pierre Luycx}
\orcid{0009-0003-8911-3553}
\affiliation{%
  \institution{University of Namur}
  \city{Namur}
  \country{Belgium}}
\email{pierre.luycx@unamur.be}

\author{Hosam Elkoulak}
\orcid{0009-0002-0043-6114}
\affiliation{%
  \institution{University of Namur}
  \city{Namur}
  \country{Belgium}}
\email{hosam.elkoulak@unamur.be}

\author{Florentin Rochet}
\orcid{0000-0001-5275-9308}
\affiliation{%
  \institution{University of Namur}
  \city{Namur}
  \country{Belgium}}
\email{florentin.rochet@unamur.be}


\renewcommand{\shortauthors}{J. Dejaeghere, L. Goffaux, P. Luycx, H. Elkoulak, and F. Rochet}

\acrodef{as}[AS]{autonomous system}
\acrodefplural{as}[ASes]{autonomous systems}

\acrodef{os}[OS]{operating system}
\acrodefplural{os}[OSes]{operating systems}

\acrodef{ixp}[IXP]{Internet exchange point}
\acrodef{cdn}[CDN]{content delivery network}

\acrodef{bgp}[BGP]{Border Gateway Protocol}

\acrodef{lts}[LTS]{long-term support}

\acrodef{gdpr}[GDPR]{General Data Protection Regulation}

\acrodef{eol}[EoL]{end-of-life}

\newcommand{\qtnone}{\emph{none }}
\newcommand{\Qtnone}{\emph{None }}

\newcommand{\qtfew}{\emph{a few }}
\newcommand{\Qtfew}{\emph{A few }}

\newcommand{\qtsome}{\emph{some }}
\newcommand{\Qtsome}{\emph{Some }}

\newcommand{\qtmany}{\emph{many }}
\newcommand{\Qtmany}{\emph{Many }}

\newcommand{\qthalf}{\emph{about half }}
\newcommand{\Qthalf}{\emph{About half }}

\newcommand{\qtmajority}{\emph{majority }}
\newcommand{\Qtmajority}{\emph{Majority }}

\newcommand{\qtmost}{\emph{most }}
\newcommand{\Qtmost}{\emph{Most }}

\newcommand{\qtaall}{\emph{almost all }}
\newcommand{\Qtaall}{\emph{Almost all }}

\newcommand{\qtall}{\emph{all }}
\newcommand{\Qtall}{\emph{All }}

\begin{abstract}
    In 2019, the Tor Project enforced an \ac{eol} policy for Tor versions, leading to the rejection of outdated relays,
    amounting to a notable fraction of consensus weight.
    While this policy aids network maintenance, reduces backporting efforts, and shortens vulnerability exposure, its
    sociotechnical implications remain unstudied.

    A user study ($N=26$) reveals that relay operators, though not universally aware of the \ac{eol} policy,
    generally view it favorably.
    Operational practices vary, occasionally excluding newly installed relays from the network.

    Network simulations, grounded in historical data, assess the policy's immediate impact on Tor clients against common
    adversaries.
    Results indicate a marginal adversarial advantage, with network churn (i.e., relays entering and exiting) exerting
    a more pronounced effect on user anonymity.
    Security metrics are introduced to evaluate relay contributions against two adversary models, enabling ranking by
    individual utility and security.
    Analysis of four exclusion rounds shows that a minority of rejected relays typically account for over 50\%
    of the security provided by all excluded relays.

    Recommendations for \ac{eol} policy implementation are proposed to mitigate potential drawbacks.
\end{abstract}
\acresetall



\maketitle
\acresetall
\section{Introduction}
\label{sec:introduction}

The Tor network~\cite{dingledine_tor_2004} is a critical privacy-enhancing technology currently
used by millions of individuals~\cite{tor-usage-imc18}.
The network is developed and open-sourced by a team largely funded through various sources~\cite{funding-tor}, run
by independent operators, and audited by a large community of academics and individuals.
The Tor Project seeks operators with intrinsic motivation to run relays in the network, as there is no monetary benefit
to running a relay.
This governance model has contributed to establishing a high level of trust, especially since the developers do not
select who may contribute by running relays.
It contrasts with other networks, such as the one created by Mullvad for its multihop
VPN~\cite{mullvad, mullvad_2025} or \mbox{Apple} Relays~\cite{apple-relay}, for which both companies actively
run relays themselves and do not open-source their code.
It also contrasts with networks such as Nym~\cite{diaz_nym_2021}, where nodes are run by various independent
operators who are incentivized with NYM tokens for their contributions.

However, over the years, the openness and independence of the Tor developers have come at a price.
The diversity of relay operators has resulted in varying rates of adoption of new Tor versions.
This has forced the developers to allow old Tor versions to continue running in the network,
which has led to various issues.
Examples of discovered attacks~\cite{evans2009practical,jansen2014sniper} by academics were eventually fixed but
remained exploitable for several years in the live network,
even on updated relays, due to backward-compatibility concerns with old versions.
The Tor developers also struggled to deploy major changes to their network, often requiring compatibility code to
interface with the behaviors of different live Tor versions.
This further delayed improvements reaching Tor clients, since those improvements would occur only if the client
happened to build circuits with the most up-to-date Tor relays.

Since 2019~\cite{goulet_blog_removing_2019}, more than 15 years after Tor's initial deployment, the developers
decided to become more restrictive toward operators.
The Tor Project redesigned the lifecycle of its software, establishing an \ac{eol} policy that involves
no feature backporting to multiple older versions and introduced a rejection deadline~\cite{tor_eol_policy_2024}.
This approach effectively strikes a compromise between the independence and freedom of operators and the requirements of the
Tor network as seen by the Tor Project.
Each major version is set to expire and be rejected from the network a few months after the next major
version is officially released.
This \ac{eol} policy led to the rejection of hundreds of relays and approximately $10\%$ of relays in the
worst event.
This rejection does not necessarily occur at regular intervals, but over the last few years it has occurred about once
a year.

Outdated relays are rejected from the network based on the Tor version they advertise to the authority servers.
When creating the network consensus, the authorities will not include rejected relays in the final consensus,
effectively excluding them from participating in the network.
This means that Tor clients will never use excluded relays.
An excluded relay can usually rejoin the network by updating to a supported version.

We aim to study the tension created by this policy change and are interested in the following questions:
\begin{rqenumerate}
    \item \label{rq1} \emph{What is the sociotechnical impact of relay rejections on the anonymity provided by Tor?}
    \item \label{rq2} \emph{How can the individual security contribution provided by a Tor relay be measured?}
\end{rqenumerate}

To answer these research questions, we make the following contributions, in order of appearance:
\begin{contribenum}
    \item \label{contrib1}
    A user study to understand how relay operators perceive the policy and to learn how they operate
    and what types of frictions are usually encountered during the lifecycles of the software used to contribute to the Tor
    network~(\cref{sec:social-aspects}).
    \item \label{contrib2}
    A quantitative evaluation of the impact of rejecting relays, using simulations based on historical data from
    the Tor network and the Internet topology~(\cref{sec:anonymity-impact}).
    \item \label{contrib3} Metrics to measure the contribution of individual Tor relays to the
    network~(\cref{sec:individual}).
\end{contribenum}

In the following sections, we present our findings and provide several contributions and recommendations to the community
with a focus on social and technical aspects.
\section{Background}
\label{sec:background}

\subsection{Tor basics}
\label{subsec:torbasics}

The Tor network~\cite{tor_spec,dingledine_tor_2004} consists of approximately \num{9500} relays and nine directory
authorities scattered across the globe and operated by volunteers.
A directory authority is a special-purpose relay that maintains a list of currently connected relays and publishes an
hourly \textit{consensus} file together with the other directory authorities.
Published consensus files contain information about all active relays in the network and are used by Tor clients to make
informed decisions about which relays to select to build circuits.

A Tor circuit typically consists of three relays: guard, middle, and exit.
Each relay in the circuit is aware only of the identity of the previous and the next relay.
Thus, the guard relay knows the IP address of the client but not the destination, and likewise, the exit relay knows the
IP address of the destination but is unaware of the source of the request.

Multiple constraints apply when selecting relays to build a circuit~\cite{tor_spec}%
\footnote{~A recent proposal suggests relaxing those restrictions.
The topic is under discussion at the time of writing.
See \shortlink{spec.torproject.org/proposals/354-relaxed-restrictions.html}.}.
The same relay cannot be selected twice in a circuit.
Additionally, any relay in the same family, administered by the same person or organization, will not be chosen as
another hop in the same path.
Moreover, it is not permitted to select more than one relay in a given network range, which defaults to \texttt{/16}
for IPv4 and \texttt{/32} for IPv6.

During path selection, when considering a relay for a specific position in a circuit, the choice is made based on the
\textit{flags} and \textit{weights} of the relay, along with positional factors called \textit{bandwidth weights}.
These criteria enable clients to compute the \textit{bandwidth} for a specific relay in a specific position.
Complete details of these load-balancing equations and their calculations are documented in related
works~\cite{rochet2017waterfilling, torweights-pets2023}.

\begin{figure}
    \centering
    \begin{overpic}[scale=.65]{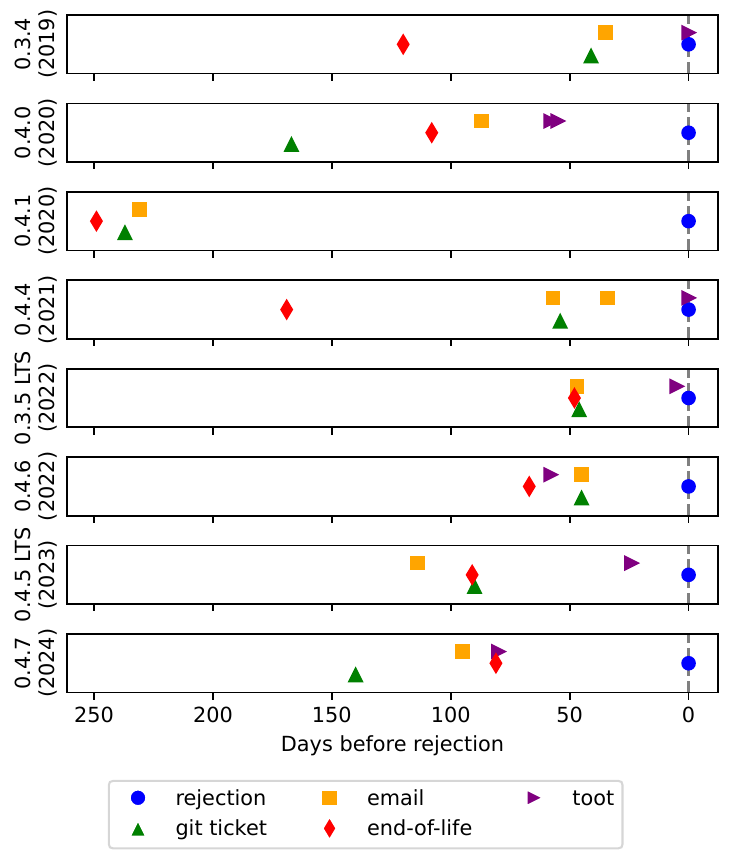}
        \boxlink{193.540674}{21.27125}{https://gitlab.torproject.org/tpo/core/team/-/wikis/NetworkTeam/CoreTorReleases}
        \boxlink{283.718928}{26.89975}{https://gitlab.torproject.org/tpo/core/tor/-/issues/31549}
        \boxlink{290.56791}{15.64275}{https://lists.torproject.org/pipermail/tor-relays/2019-September/017711.html}
        \boxlink{330.520301}{13.64275}{https://mastodon.social/@torproject/102927631601760864}
        \boxlink{330.520301}{21.27125}{https://gitlab.torproject.org/tpo/network-health/team/-/wikis/Relay-EOL-policy\#septemberoctober-2019}

        \boxlink{139.890319}{69.34225}{https://gitlab.torproject.org/tpo/core/tor/-/issues/32672}
        \boxlink{207.238636}{63.71375}{https://gitlab.torproject.org/tpo/core/team/-/wikis/NetworkTeam/CoreTorReleases}
        \boxlink{231.210071}{58.08525}{https://lists.torproject.org/pipermail/tor-relays/2020-February/018185.html}
        \boxlink{262.313481}{58.08525}{https://mastodon.social/@torproject/103874328061744976}
        \boxlink{267.737972}{58.08525}{https://mastodon.social/@torproject/103890902321687937}
        \boxlink{330.520301}{63.71375}{https://lists.torproject.org/pipermail/tor-relays/2020-May/018582.html}

        \boxlink{46.287574}{106.15625}{https://gitlab.torproject.org/tpo/core/team/-/wikis/NetworkTeam/CoreTorReleases}
        \boxlink{59.985537}{111.78475}{https://gitlab.torproject.org/tpo/core/tor/-/issues/34357}
        \boxlink{66.834518}{100.52775}{https://lists.torproject.org/pipermail/tor-relays/2020-June/018594.html}
        \boxlink{330.520301}{106.15625}{https://metrics.torproject.org/versions.html?start=2021-01-01\&end=2021-01-30}

        \boxlink{137.607326}{148.59875}{https://gitlab.torproject.org/tpo/core/team/-/wikis/NetworkTeam/CoreTorReleases}
        \boxlink{265.454978}{142.97025}{https://lists.torproject.org/pipermail/tor-relays/2021-October/019862.html}
        \boxlink{268.879469}{154.22725}{https://gitlab.torproject.org/tpo/core/tor/-/issues/40480}
        \boxlink{291.709407}{142.97025}{https://lists.torproject.org/pipermail/tor-relays/2021-October/019933.html}
        \boxlink{330.520301}{140.97025}{https://mastodon.social/@torproject/107372493082377481}
        \boxlink{330.520301}{148.59875}{https://gitlab.torproject.org/tpo/network-health/team/-/wikis/Relay-EOL-policy\#septemberoctober-2021}

        \boxlink{276.869947}{184.41275}{https://lists.torproject.org/pipermail/tor-relays/2022-February/020289.html}
        \boxlink{274.72845}{191.04125}{https://gitlab.torproject.org/tpo/core/team/-/wikis/NetworkTeam/CoreTorReleases}
        \boxlink{278.011444}{196.66975}{https://gitlab.torproject.org/tpo/core/tor/-/issues/40559}
        \boxlink{324.812817}{185.41275}{https://mastodon.social/@torproject/107967484645149143}
        \boxlink{330.520301}{191.04125}{https://gitlab.torproject.org/tpo/network-health/team/-/wikis/Relay-EOL-policy\#februarymarch-2022}

        \boxlink{254.040009}{233.48375}{https://gitlab.torproject.org/tpo/core/team/-/wikis/NetworkTeam/CoreTorReleases}
        \boxlink{264.313481}{227.85525}{https://mastodon.social/@torproject/108799351376748467}
        \boxlink{279.152941}{227.85525}{https://lists.torproject.org/pipermail/tor-relays/2022-August/020765.html}
        \boxlink{279.152941}{239.11225}{https://gitlab.torproject.org/tpo/core/tor/-/issues/40664}
        \boxlink{330.520301}{233.48375}{https://gitlab.torproject.org/tpo/network-health/team/-/wikis/Relay-EOL-policy\#augustseptember-2022}

        \boxlink{200.389655}{270.29775}{https://lists.torproject.org/pipermail/tor-relays/2023-January/020985.html}
        \boxlink{226.644084}{274.92625}{https://gitlab.torproject.org/tpo/core/team/-/wikis/NetworkTeam/CoreTorReleases}
        \boxlink{227.78558}{282.55475}{https://gitlab.torproject.org/tpo/core/tor/-/issues/40760}
        \boxlink{303.124376}{270.29775}{https://mastodon.social/@torproject/110256009579078954}
        \boxlink{330.520301}{275.92625}{https://gitlab.torproject.org/tpo/network-health/team/-/wikis/Relay-EOL-policy\#marchapril-2023}

        \boxlink{170.710736}{323.99725}{https://gitlab.torproject.org/tpo/core/tor/-/issues/40896}
        \boxlink{222.078096}{312.74025}{https://lists.torproject.org/pipermail/tor-relays/2024-January/021470.html}
        \boxlink{239.200549}{311.74025}{https://mastodon.social/@torproject/111858473454854394}
        \boxlink{238.059053}{319.36875}{https://gitlab.torproject.org/tpo/core/team/-/wikis/NetworkTeam/CoreTorReleases}
        \boxlink{330.520301}{318.36875}{https://metrics.torproject.org/versions.html?start=2024-04-01\&end=2024-05-01}
    \end{overpic}
    \caption{An overview of the timeline of events preceding the exclusion of outdated relays from the Tor network.}
    \label{fig:back:timeline}
\end{figure}

\subsection{Updating Tor relays}
\label{subsec:updating-tor-relays}

Before 2019, limited restrictions were enforced on the Tor versions accepted as relays in the network.
Because of this, by late 2019 the Tor network comprised relays running 85 different Tor versions, with the
oldest version released in December 2013~\cite{goulet_blog_removing_2019}.
Some of those versions were still supported by the Tor Project, and some were no longer supported.

This situation posed two main challenges for the Tor Project.
First, it was maintaining five version series (i.e.,\ major versions) to provide fixes for major stability issues,
security vulnerabilities, and portability regressions~\cite{goulet_blog_removing_2019}.
Second, operating the network with many versions made it more difficult to deploy certain security fixes and new features.
Changes were generally available only for the supported versions, but those changes needed to account for the
unsupported versions still operating in the network.

Because of these challenges, the Tor Project decided to stop accepting relays running unsupported versions.
The project began contacting operators running such versions, asking them to update their relays to continue participating
in the Tor network.

Currently, the Tor Project handles the process of rejecting \ac{eol} versions using public announcements (via its
mailing list~\cite{mailing_tor_relays} and its social media accounts~\cite{fedi_tor_2024, twitter_tor_2020}) and
direct emails to relay operators who provided contact information when setting up their relays.
Since the first rejection of relays in 2019, the Tor Project has proceeded as follows: once a version reaches
\ac{eol}, the Network Health team sends an email to the Tor relay mailing list and opens a git ticket to track
changes to be made in the source code to remove the rejected versions effectively.
The time elapsed between a version becoming \ac{eol} and its actual exclusion was significantly longer in the first
exclusion rounds.
For the last exclusion round, the notification to the mailing list was sent before the version became \ac{eol}.
\cref{fig:back:timeline} provides an overview of the different events leading to the exclusion of eight Tor versions.
\section{Social aspects}
\label{sec:social-aspects}

To address the social component of~\cref{rq1}, we collected feedback from volunteer relay operators regarding their
experiences as Tor relay operators.
Our objective is to determine whether relay operators are aware of the \ac{eol} policy enforced by the Tor Project and
how they perceive the value of this policy.
We also examine how relay operators keep their relays up to date (if they do) and how they react when, or if, their
relays are rejected from the network.

\subsection{Method}
\label{subsec:method}

To investigate how Tor relay operators manage their relays and their views on the \ac{eol} policy and exclusion cycle,
we conduct a qualitative study of current and past operators.
Following~\citet{lazar2017research}, our study has two parts.

First, we construct an interview guide for semi-structured interviews and conduct a pilot interview with a
colleague running a Tor relay to identify areas of interest not covered by our initial interview guide.
Based on this pilot interview, we augment the guide with additional topics of interest.
As the colleague is aware of our research direction and thesis, we do not include the pilot interview in the
results.

Second, we contact Tor relay operators via the tor-relays mailing list and via printed signs at
FOSDEM\footnote{~\shortlink{fosdem.org/2026/}} in Brussels.
Interested Tor relay operators take the interview in one of the following settings: in person during FOSDEM, via an
online voice call, via a synchronous text chat, or via asynchronous emails.
In-person interviews and online voice calls are audio-recorded after obtaining consent from the participant.
For text-based interviews, we save the transcript for analysis.
We discuss ethical considerations in more detail in~\cref{sec:ethics}.
The recruitment process and the interview guide are presented in \cref{app:interview-process}.

We acknowledge that there might be significant selection bias in how we recruit participants: we are interested in
how relay operators manage their relays with a focus on the \ac{eol} policy.
Finding participants via the mailing list and at FOSDEM may not yield a representative set,
as the least dedicated relay operators are probably not following the mailing list or attending FOSDEM\@.
For this reason, we adhere to a qualitative analysis and refrain from conducting a quantitative analysis (e.g., via
an online form with closed questions) to avoid amplifying the selection bias.

\subsection{Analysis}
\label{subsec:analysis}

Between January 31 and February 8, 2026, we conducted 26 interviews.
We conducted the interviews as follows: 6 in-person interviews at FOSDEM, 12 online voice calls, 7 synchronous text
interviews, and 1 asynchronous text interview.
The same researcher conducted all interviews.
All interviews but two were conducted in English, which is not the native language of the researcher and often not
the native language of the participant either.
The two remaining interviews were conducted in the native language of both the interviewer and the participant.

Using transcriptions of the audio interviews and saved transcripts of the text interviews, we develop a codebook by
having two coders openly and independently code a set of three interviews.
From the codes created by the two coders, we build a common codebook used to code all interviews.
We augment this common codebook when new codes emerge.
The resulting common codebook has \num{137} codes with \num{1511} coded segments in total.

Both coders regularly check the coded interviews to resolve potential points of disagreement and to align their
interpretations of the data.
We compute Cohen's kappa~\cite{cohen_coefficient_1960,lazar2017research} score per code at the interview level.
The average score for the two coders is \num{0.78} (SD \num{0.23}), indicating satisfactory agreement~\cite{lazar2017research}.

\begin{figure}
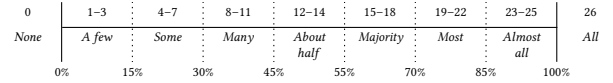

    \Tiny
    \begin{tabularx}{0.4725\textwidth}
        {>{\centering\arraybackslash}X|>{\centering\arraybackslash}X;{0.5pt/1.5pt}>{\centering\arraybackslash}X;
                {0.5pt/1.5pt}>{\centering\arraybackslash}X;{0.5pt/1.5pt}>{\centering\arraybackslash}X;{0.5pt/1.5pt}
                >{\centering\arraybackslash}X;{0.5pt/1.5pt}>{\centering\arraybackslash}X;{0.5pt/1.5pt}
                >{\centering\arraybackslash}X|>{\centering\arraybackslash}X}
\rule{0pt}{2.5ex} 0 & 1--3 & 4--7 & 8--11 & 12--14 & 15--18 & 19--22 & 23--25 & 26 \\[1ex] \cline{2-8}
\rule{0pt}{2.5ex} \Qtnone & \Qtfew & \Qtsome & \Qtmany & \Qthalf & \Qtmajority & \Qtmost & \Qtaall & \Qtall \\[1ex]
    \end{tabularx}

    \begin{tabularx}{0.42\textwidth}
        {>{\centering\arraybackslash}X>{\centering\arraybackslash}X>{\centering\arraybackslash}X
                >{\centering\arraybackslash}X>{\centering\arraybackslash}X>{\centering\arraybackslash}X
                >{\centering\arraybackslash}X>{\centering\arraybackslash}X}
 0\%   &  15\%     & 30\%     & 45\%     & 55\%           &  70\%        & 85\%     & 100\%
    \end{tabularx}
    \caption{Quantifiers and percentages used to present the qualitative results. Each quantifier always refers to
    the same number of participants throughout the section. \emph{Emphasis} is used when the quantifier appears in
    the text to remind the reader that it carries a specific meaning.}
    \label{fig:quantifiers}
\end{figure}

Similar to previous work~\cite{habib_its_2020,emami-naeini_exploring_2019,zhang_how_2022,usman_distrust_2023,klivan_everyone_2024,krause_thats_2025},
we report the prevalence of our findings using the quantifiers presented in \cref{fig:quantifiers} instead of
reporting exact counts.

\subsection{Qualitative results}
\label{subsec:qualitative-results}

In total, participants manage \num{1379} relays located in \num{36} distinct countries.
On average, participants manage \num{53.1} relays (SD \num{148.4}) across \num{2.6} countries (SD \num{2.1}).
The \qtmajority of participants manage 5 relays or fewer; \qtsome participants manage a few dozen relays, and \qtsome
participants manage hundreds of relays.
\Qtmany participants have managed relays for less than 5 years, and the \qtmajority of participants have managed relays
for 5 years or more.
The \qtmajority of participants are solo operators, and \qtmany are part of 8 distinct groups or organizations.
We choose not to give more precise profiles to limit information exposure and potential reidentification.

Although our participant sample might be biased toward committed relay operators, \qtsome participants are not
aware of the \ac{eol} policy.
The other participants are at least vaguely aware of the policy.

\subsubsection{Update workflow}

The \qtmajority update Tor manually: the process needs to be started by a human operator.
However, the manual update is often triggered by a regular reminder or a message on the Tor Project mailing list
announcing a new release.
Among the participants stating that they manually update Tor, we find people operating large numbers of relays who
still prefer to update manually.
For example: \itwquote{So we only update when it's needed, for example, when a security patch is relevant to our
situation or when the Tor Project actually marks a previous version as \ac{eol} or dangerous or bugged.}~(P12)
Participants mention various motivations to manually update:
for \qtfew participants, there are no automatic ways of updating on their \ac{os};
\qtfew participants want to manually check the changelog;
\qtsome do not need their update process to scale up;
\qtsome feel that the process is simpler, simple enough, or familiar;
\qtfew want to evaluate whether the relay reputation hit is worth taking
or want to test the update on a few relays before applying it to all relays.
Finally, one participant mentions that they never update their relay:
\itwquote{as an extremely casual relay operator, I mostly don't [update my relays]}~(P25).
Another participant performing manual updates says:
\itwquote{I honestly forget I run Tor nodes unless something makes me think about them}~(P10).

Among participants updating automatically, several also manually update when requested by the Tor Project
if their update schedule is infrequent.
\itwquote
{Packer rebuilds the cloud images at least once per month on a schedule, but we can trigger immediate rebuilds in
case of CVEs or other notices that require a quicker turnaround.}~(P16)
Motivations to automatically update include
familiarity with this process,
supporting the Tor developers,
applying the latest security patches, and
reducing maintenance burden.
It is important to note that operators who manually update their relays are not necessarily slower to update than those with
automatic updates: automatic updates might run on a monthly schedule, whereas manual updates can be performed upon
notification on the Tor Project mailing list, for example.
\itwquote{I'm on the announcement mailing list, so when I see that there's a new release then I usually update it
certainly within a week, maybe within a few days.}~(P08)

\subsubsection{Tor package repository}

\Qtmany participants report that they use the default package repository of their distribution for Tor.
\Qtmost participants use Debian and/or Ubuntu.
\Qtsome participants use other Linux distributions, such as Alpine or NixOS, for instance.
There are also \qtfew participants who use FreeBSD and/or OpenBSD with the goal of increasing \ac{os} diversity on the
network.
It is not unusual for \qtsome participants to run multiple distributions and/or \acp{os}.

\begin{figure}
    \centering
    \includegraphics[width=\linewidth]{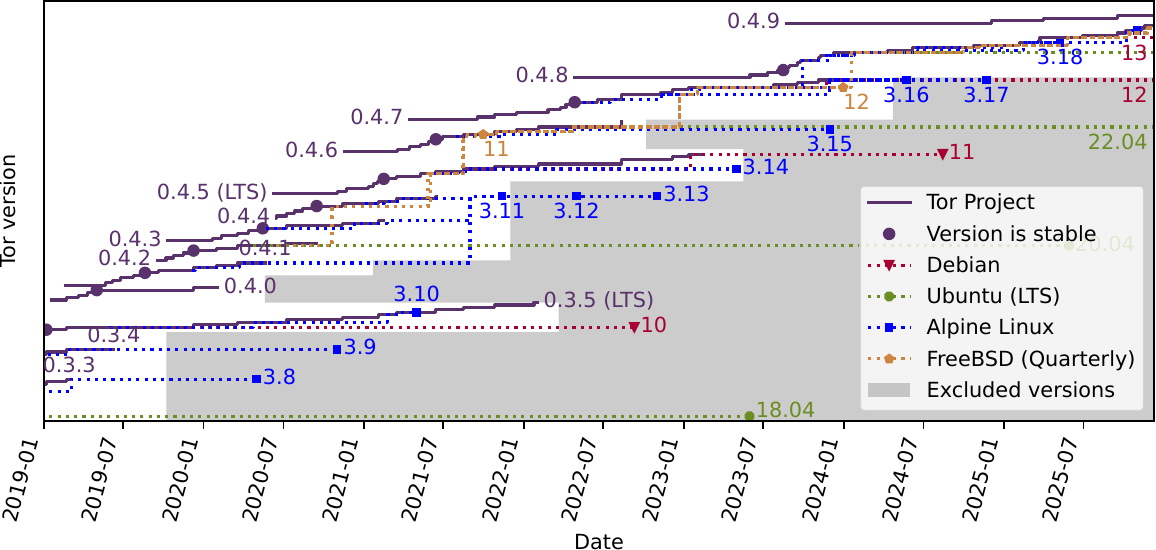}
    \caption{Tor binary packages provided by different \ac{os} software repositories (dotted lines)
        compared with the Tor Project's official source code releases (full lines). Dotted lines stop when
        an \ac{os} version is no longer supported,
        annotated by a marker. Full lines stop when a Tor version series reaches
        \ac{eol}.}
    \label{fig:tor-pkg-version}
\end{figure}

We analyze the version of Tor shipped by different \acp{os} in relation to the Tor versions accepted in the network
since 2019.
\cref{fig:tor-pkg-version} provides a visual representation of Tor versions shipped by \ac{os} repositories.
Whether we look from the Tor Project's point of view or from package maintainers' point of view, multiple versions
are maintained simultaneously in both cases.
The grayed part represents versions rejected from the network.
We notice that Debian generally ends up shipping a Tor version that is not accepted anymore toward the end of its
lifetime, even if the Tor version it ships is marked as \ac{lts}.
The same applies to Ubuntu \ac{lts} versions.
In addition, however, Ubuntu \ac{lts} does not generally ship an \ac{lts} version of Tor.
For example, if Ubuntu 22.04 (\ac{lts}) shipped Tor 0.4.5 (\ac{lts}) instead of 0.4.6, it would still have been
excluded, but 5 months later.
Even \acp{os} with shorter release cycles, such as Alpine Linux or FreeBSD, sometimes ship outdated Tor versions.
This is due to multiple release life cycles having different schedules, philosophies, and maintainers.

\Ac{os} release cycles can affect the ability of volunteers to participate in the Tor network.
Some operators are not always aware of this: \itwquote{You might know that Debian has some very old packages in it.
It took me a while before I learned that actually if you just enable the ORPort on Debian Tor, it's rejected by the
network because it's too old.}~(P25)
Even if aware, one relay operator using Debian with the default Tor package mentions that \itwquote{The relay was still
seeing traffic at the time, so I thought I'll just let it be and it continued to receive and process traffic, so that
was fine.}~(P03)

We note that the Tor Project is aware of the problem.
In their installation guides, they recommend using their own package repositories for Debian, Ubuntu, RHEL,
and Fedora~\cite{torproject_installing}.
Concerning FreeBSD, they recommend using \say{latest} package releases instead of the default \say{quarterly} option
in order to get faster Tor updates~\cite{torproject_freebsd}.

Package maintainers providing fast updates help operators run the latest version without too much burden:
\itwquote{[FreeBSD and OpenBSD] both have high-quality Tor packages available, because there is Yuri who maintains
high-quality packages for FreeBSD and OpenBSD. So we always have them fast, all the updates.}~(P12)
However, they recognize that this burden is instead on the package maintainer's shoulders, who need free time to provide
updates, and that it could be a single point of failure if the maintainer is absent or if the package repository is
down.

\subsubsection{Opinion about \ac{eol} policy}

During the interviews, we provide a brief description of the \ac{eol} policy to the participants.
We then ask for their opinions about this policy.
\Qtmost participants are aware of the policy, and \qtmost are positive about it.
This positive reaction is generally the first point they express when asked for their opinion.
\Qtsome have either no strong opinion or understand the rationale without an explicitly positive sentiment in their
wording.
\Qtaall participants support the policy for the security benefits they perceive: new releases often fix potential
vulnerabilities or introduce new security-related features.
For \qtsome participants, updating is also important to help the developers deploy new features and remove the
burden of maintaining multiple version series concurrently.

While participants are generally positive regarding the \ac{eol} policy, some nuanced aspects are mentioned.
According to \qtfew participants, the policy adds a maintenance burden on operators, who must keep systems up to date
and track which version is about to be rejected.
This burden is generally considered reasonable by the operators who mention it. 
However, one participant thinks that, while it is normal to expect this from \itwquote{middle- to large-scale operators},
it can discourage small operators: \itwquote{There are also many relay operators that run like one relay or two relays
[...], they didn't sign up for having monthly maintenance.}~(P12)
Another participant thinks the burden is reasonable only if operators are notified one or two months in advance (P18).
\Qtsome participants are concerned that the policy might reduce the diversity of relays in the network.
Indeed, less committed operators might run relays in otherwise uncovered areas and be removed from the network if
they do not update frequently enough.
We note that, except for two of them, these nuanced operators have managed relays for five years or more.
Despite the nuances, participants value the Tor Project and its community for their voluntary support.

Finally, based on participants' wording, it was not always clear whether they could distinguish between
the \ac{eol} policy and the \say{not recommended} flag that was recently applied to some 0.4.8.x versions.
Being clearer about the \ac{eol} policy when newcomers set up a Tor relay is suggested by \qtfew participants.
Clear and proactive communication from the Tor Project is important to \qtsome participants; they appreciate
emails from the Tor Project notifying them that a relay they operate is down or has an issue.
This was also discussed on the mailing list~\cite{weaver_accelerated_2025,nusenu_tor-relays_2026}.

\subsubsection{Reaction if rejected}

As mentioned, the participants we interviewed are likely among the most committed relay operators.
Only two participants are certain that one of their relays was excluded from the network.
Both of them updated the Tor package to continue participating in the network.

\Qtsome participants would be annoyed, \qtsome embarrassed, or \qtsome surprised at first if their relay were
excluded from the network.
\Qtsome would feel as though they had failed in their duty as relay operators.
Given the results presented in the previous sections, multiple operators acknowledge that they would understand why
their relay was rejected.
Most importantly, \qthalf of the participants mention that they would update their relay if they learned that
it was excluded.
\Qtmany also mention that it would prompt them to revise their update process.

\begin{figure}
    \centering
    \includegraphics[width=\linewidth]{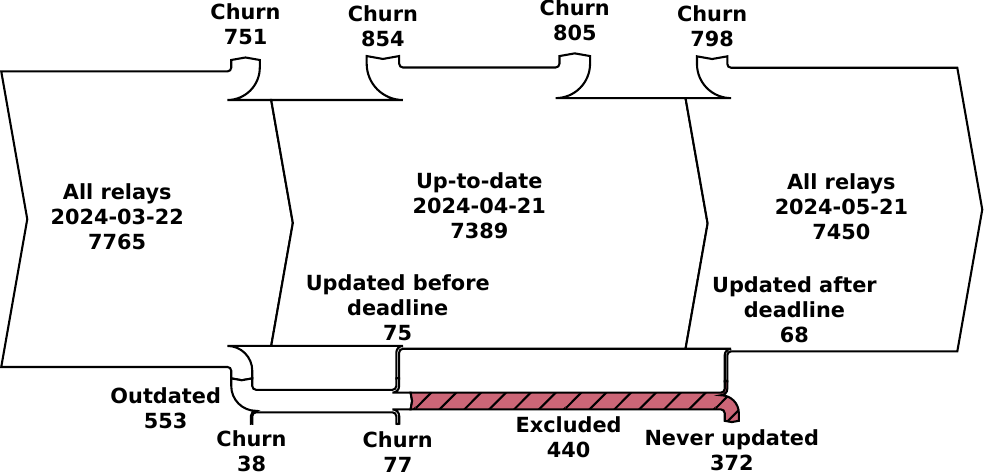}
    \caption{Flow of Tor relays before and after the 2024-04 exclusion (< 0.4.8.0-alpha-dev).
    \emph{Churn} flows represent relays entering and leaving the network, unrelated to the \ac{eol} policy.
    \emph{Outdated} relays run a version that is older than the one accepted on 2024-04-21; if they update, they join the
    \emph{Up-to-date} flow, otherwise, they fall into the \emph{Excluded} flow.}
    \label{fig:tor-sankey}
\end{figure}

In practice, however, most excluded relays never return.
To obtain an actual number, we sample historical consensuses from the directory authorities~\cite{collector} at three
different dates: one month before, just before, and one month after an exclusion date.
\cref{fig:tor-sankey} presents our results regarding the April 2024 exclusion.
Out of the \num{7765} relays in the network, \num{553} had a to-be-excluded Tor version one month before.
\num{75} of them updated before the exclusion deadline, while 440 did not and were excluded.
Finally, by looking one month later, we observe that \num{68} (15\%) excluded relays returned with the same fingerprint,
while \num{372} (85\%) excluded relays did not return (or not with the same fingerprint, at least).
In addition, the chart shows that several hundred up-to-date relays enter and leave the network
over the course of one month.
This flow is what we call the natural churn of relays.

\subsubsection{Key takeaways}

There is no single way to operate relays in the Tor community.
For various reasons, the \qtmajority of operators prefer manual updates, whereas others prefer automatic updates of the
Tor daemon.
Although the Tor Project recommends against using default package repositories for well-known
\acp{os}~\cite{torproject_installing}, we encountered operators who use them.
The differing lifecycles of \ac{os} distributions may result in fresh relays from new contributors being
excluded immediately.
\Qtsome participants are aware of the \ac{eol} policy.
Yet, even within our sample of dedicated operators, \qtmost were only vaguely aware.
Others became aware of it only after their relay was actually excluded.
The policy is widely supported, primarily for improving network security, enabling faster feature deployment by Tor
developers, and reducing maintenance by limiting parallel versions.
However, experienced operators highlight concerns about reduced diversity, unclear policy communication, and the
impact of announcement timing.

In the next sections, we define Tor's threat model and adversaries and analyze how the rejection of relays due to the
\ac{eol} policy impacts its security (i.e.\ its anonymity).
This direction is supported by the results of the quantitative analysis presented: participants generally support the
\ac{eol} policy and cite the security of the network as a main benefit.
We also analyze the impact of the hundreds of relays entering and leaving the network, irrespective of the \ac{eol} policy.

\section{Threat model}
\label{sec:threat-model}

Tor's threat model aims to defend and provide a degree of anonymity to users, on a best-effort basis, against
non-global adversaries.
Prior literature has developed several deanonymization techniques employing local adversaries to
identify Tor users~\cite{murdoch2005low, juarez2014critical, rochet_dropping_2018, kohls2018digestor}.
These techniques are generally feasible when an adversary can intercept communication on both sides of a Tor circuit.
The adversary can then correlate traffic between the client and the guard with traffic between the exit and the final
destination~\cite{murdoch2005low}.
Such attacks are called traffic confirmation attacks: they reveal that a given client connected to a given
destination address.
In this threat model, adversaries include relays within the Tor network itself; an adversary controlling a portion of
end paths could position itself as both guard and exit relays in circuits to deanonymize users~\cite{rochet_dropping_2018}.
Other potential adversaries include \acp{as}~\cite{feamster2004location}
that handle Internet traffic passing through their networks and are therefore capable of correlating traffic at each
end of a circuit.
However, these are not the only entities capable of observing Tor user traffic.
Some studies~\cite{murdoch2007sampled,juen2012protecting, johnson_torps_2013} also suggest considering \acp{ixp}
as potential threats.

Similar to~\citet{barton_denasa_2016, sun2017counter, claps}, we categorize these adversaries into two classes.
The first class is the relay adversary, which controls a subset of Tor relays to deanonymize users.
The second class is the network-level adversary, namely an \ac{as} or \ac{ixp},
capable of intercepting communications on both sides of a circuit when traffic passes through them.
We consider such entities suspect if they are able to observe both ends of Tor circuits.
\section{Exclusion anonymity impact}
\label{sec:anonymity-impact}
To study the technical anonymity impact that relay rejections had on Tor users (\cref{rq1}), we created an updated version
of TorPS (\citeauthor{johnson_torps_2013}, \citeyear{johnson_torps_2013}~\cite{johnson_torps_2013}) to match the path
selection algorithm in effect during the exclusion periods.
This tool evaluates, via a Monte Carlo method, the probability of compromise that a typical Tor client faces while using
the Tor network.
This probability is cumulative as time advances: the more a Tor client uses the network, the more likely it is to be
compromised.
We measure this cumulative probability over time for a typical Tor client.
We evaluate the impact that relay rejections had on Tor clients during a timeframe spanning 14 days before to 14 days
after an exclusion date to gauge the risk created by rejecting honest relays.
We assume that potentially malicious relays are up to date or misreport their versions and therefore would not be
rejected.
This is a reasonable assumption: attacking the network already takes high effort, whereas announcing a fake version or updating a relay takes little effort.

The exclusions considered are the following: 2021-12-01, 2022-03-21, 2022-10-07, and 2023-05-17~\cite{tor_eol_policy_2024}.
These exclusion rounds are those for which the Tor Project shared the list of excluded relays.
We also use the algorithm from~\citet{qiu_as_2006} to infer the \ac{as}
path between the client and the guard and between the exit and the destination.
\citet{qiu_as_2006} use \ac{bgp} data to learn about existing paths and neighboring \acp{as} on the Internet.
When a path is announced via \ac{bgp}, \citet{qiu_as_2006} save it as what they call a \emph{sure path}.
During inference, paths might be extended using neighboring \acp{as}, and the number of hops added to a path using
neighboring \acp{as} is called the \emph{unsure length}.
When inferring a path from a given \ac{as} to a destination prefix, we use the \emph{Least Uncertainty First}
strategy from \citet{qiu_as_2006}: the inference procedure returns paths with the smallest \emph{unsure length},
ordered by the probability that the path is used on the Internet.
This probability is derived from the number of times the path has been observed in \ac{bgp} data.
We always take the first result as the inferred path to ensure reproducible results across executions.

Our goal is to measure the impact of the exclusions on Tor clients' anonymity.
To this end, we consider
\begin{inparaenum}[(i)]
    \item a network adversary compromising some \acp{as} (\cref{subsec:as-adversary}),
    \item a network adversary compromising some \acp{ixp} (\cref{subsec:ixp-adversary}), and
    \item a relay adversary adding malicious relays to the Tor network (\cref{subsec:relay-adversary}).
\end{inparaenum}
Following \citet{johnson_torps_2013}, for each exclusion and adversary type, we compute the probability of compromise
over time.
A client is considered compromised as soon as an adversary compromises one of its circuits.

Similar to previous
literature~\cite{claps,wan2019guard, barton_denasa_2016, sun2017counter, dinh2020scaling, rochet2022towards}, we use
data from multiple sources to reproduce the Tor network state during past exclusions.
The data used are time-dependent whenever available; that is, the past is reconstructed as
faithfully as the available data allow.

In total, four sets of simulations are run (one set per exclusion) with the following parameters:
\begin{itemize}
    \item 10,000 Tor clients whose locations are computed based on Tor Metrics~\cite{tor_metrics_users} and on the \ac{as} ranking provided by CAIDA~\cite{caida_as_rank};
    \item the 100 most popular destinations, taken from Tranco~\cite{le_pochat_tranco_2019}, which provides a ranking of
    popular domains by aggregating the rankings of other providers over a period of 30 days;
    \item consensus documents from the Tor network~\cite{tor_sources} to model the Tor topology;
    \item \ac{bgp} and \ac{ixp} data to model the topology of the Internet, from \citet{routeviews_2024} and the
    \citet{caida_ixps}, respectively.
\end{itemize}
For each set, two simulations are run: one with the historical data~\cite{tor_sources} and one with the
historical data augmented with the relays that were scheduled for exclusion, effectively simulating the network as if the
exclusion had not occurred.
The difference between the two results represents the impact of the exclusion.
The results we obtain from the simulations may differ if we decide to change the above parameters.
However, this setup reflects the general trend of the evolution of user anonymity.
Moreover, the differences between two experiments using the same setup are in any case worth comparing.

\subsection{\ac{as} adversary}
\label{subsec:as-adversary}

\begin{figure*}
    \centering
    \begin{subfigure}[b]{0.24\textwidth}
        \centering
        \includegraphics[width=.9\textwidth]{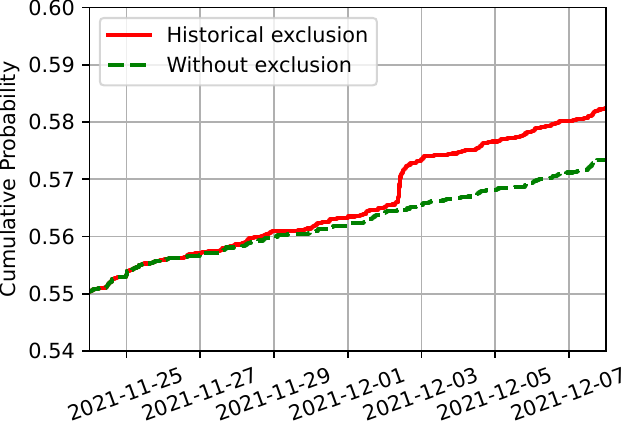}
        \caption{Exclusion of 2021-12-01.}
        \label{fig:results:net-as-adv:2021-12}
    \end{subfigure}
    \begin{subfigure}[b]{0.24\textwidth}
        \centering
        \includegraphics[width=.9\textwidth]{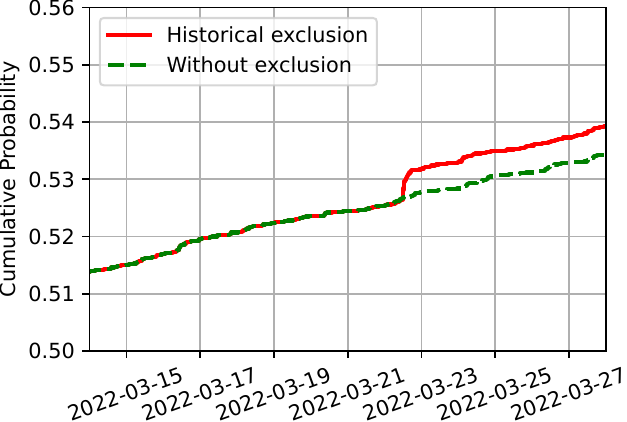}
        \caption{Exclusion of 2022-03-21.}
        \label{fig:results:net-as-adv:2022-03}
    \end{subfigure}
    \begin{subfigure}[b]{0.24\textwidth}
        \centering
        \includegraphics[width=.9\textwidth]{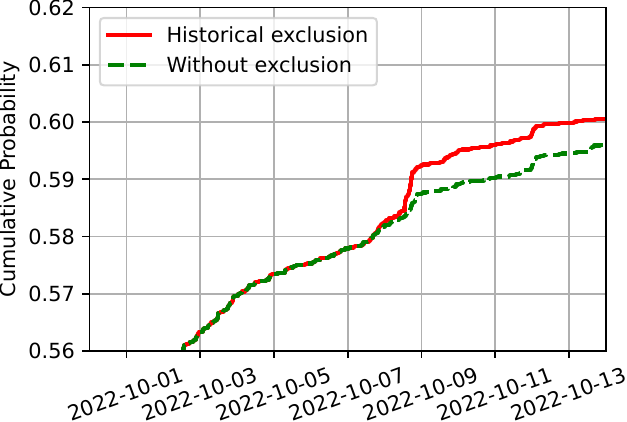}
        \caption{Exclusion of 2022-10-07.}
        \label{fig:results:net-as-adv:2022-10}
    \end{subfigure}
    \begin{subfigure}[b]{0.24\textwidth}
        \centering
        \includegraphics[width=.9\textwidth]{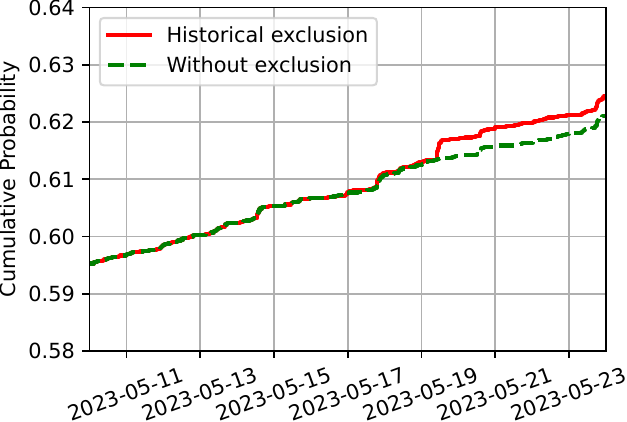}
        \caption{Exclusion of 2023-05-17.}
        \label{fig:results:net-as-adv:2023-05}
    \end{subfigure}
    \caption{Probability of path compromise with top-2 suspicious \acp{as}.
    The gap between the red and green lines shows the advantage the exclusion gave the suspicious
    \acp{as} -- roughly a few days of running the network without excluding relays.}
    \label{fig:results:net-as-adv}
\end{figure*}

In the \ac{as} adversary setting, the two most frequent
\acp{as} (i.e., the two \acp{as} that appear most frequently on both ends of the circuit) are considered suspect
\acp{as} but non-colluding.
Similar to DeNASA's approach~\cite{barton_denasa_2016}, any \ac{as} appearing at both ends of a Tor circuit is considered suspect;
however, a few \acp{as} appear significantly more frequently.
These two \acp{as} are treated as suspect \acp{as}, and the evaluation is conducted against them.
An evaluation against all of them could also be performed, which would shift the base probability closer to one.
This choice is motivated by the marked difference from the first two \acp{as}
(\cref{appendix:as-adversaries-ranking}, \cref{fig:sus-ases}).

\Cref{fig:results:net-as-adv}
shows the cumulative distribution function of compromised clients over time for four recent exclusions, based on our
simulations.
On each graph, the solid red line represents the historical data: when the exclusion occurs, some clients are impacted
and observe their guard go offline.
Those clients must select a new guard.
As the new guard is located elsewhere, migrating clients may now have a suspicious \ac{as} on the path between them and their guard.
As they connect to many destinations over time, it is also likely that the same suspicious \ac{as} is on the path
between the exit relay and the destination.
For this reason, a number of clients generally become compromised a few hours after the exclusion.

The dotted green line on each graph represents a scenario in which the exclusion does not occur.
In this case, clients do not observe their guard go offline, and they can continue using it without selecting a new
guard and potentially using a compromised path.

A general slope of the cumulative distribution function is also observed, and analysis of the data reveals that the
natural churn of relays in the network (observed earlier in \cref{fig:tor-sankey}) is the main factor.
Another factor is the guard-flag assignment logic for some low-bandwidth relays.
The data reveal that some low-bandwidth relays gain and lose their guard flag due to variation in their bandwidth as
measured by the authorities.
This instability could be mitigated by slightly increasing the bandwidth bar to gain the guard flag and reducing the bar
to lose it.

It is noteworthy that around 1\% of additional clients were compromised during the 2021 exclusion, whereas
only about 0.30\% were compromised in 2023.
This decrease over time can be attributed to two main factors.
First, relay operators may have become accustomed to updating their relays regularly; otherwise, their relays are rejected from
the network.
Second, the Tor Project gained insights from previous exclusions and improved the workflow over the years to
keep relay operators up to date with the policy.
This decrease is also evident in mailing list communications from the Tor Project: in their email regarding the
2019 exclusion round, they stated that the excluded relays accounted for 13.25\% of the consensus weight (computed
approximately one month before the exclusion)~\cite{nusenu_tor-relays_2019}.
For the 2023 exclusion round, the Tor Project stated that the excluded relays accounted for \say{roughly 7\% of the
advertised bandwidth of the network} (computed five months before the exclusion)~\cite{koppen_tor-relays_2023}.

\begin{figure*}
    \centering
    \begin{subfigure}[b]{0.24\textwidth}
        \centering
        \includegraphics[width=.9\textwidth]{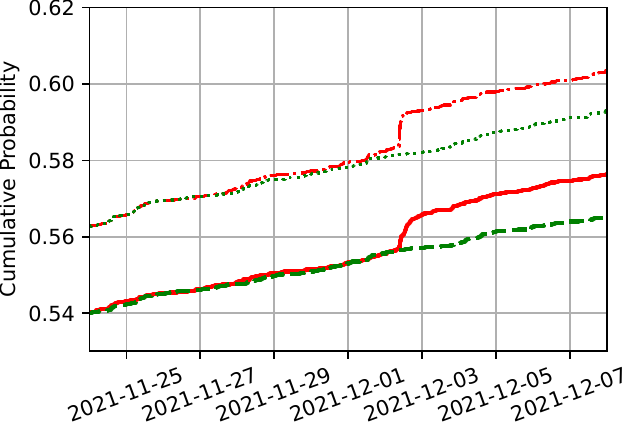}
        \caption{Exclusion of 2021-12-01.}
        \label{fig:net-org-adv:2021-12}
    \end{subfigure}
    \begin{subfigure}[b]{0.24\textwidth}
        \centering
        \includegraphics[width=.9\textwidth]{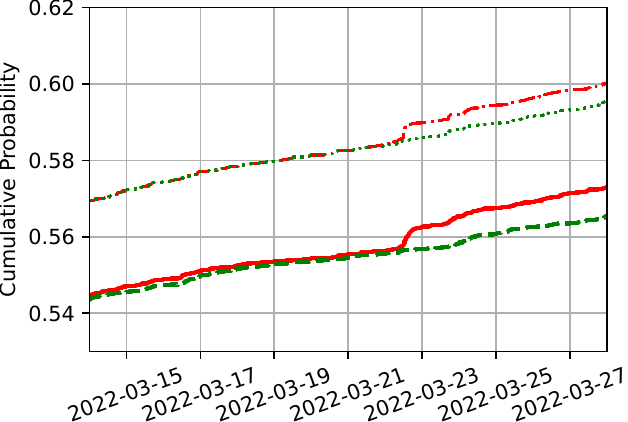}
        \caption{Exclusion of 2022-03-21.}
        \label{fig:net-org-adv:2022-03}
    \end{subfigure}
    \begin{subfigure}[b]{0.24\textwidth}
        \centering
        \includegraphics[width=.9\textwidth]{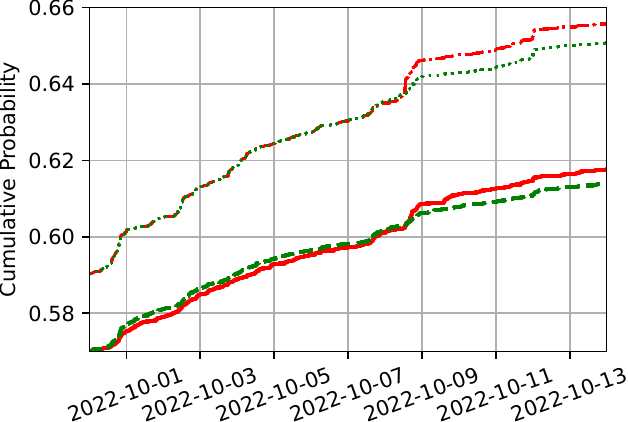}
        \caption{Exclusion of 2022-10-07.}
        \label{fig:net-org-adv:2022-10}
    \end{subfigure}
    \begin{subfigure}[b]{0.24\textwidth}
        \centering
        \includegraphics[width=.9\textwidth]{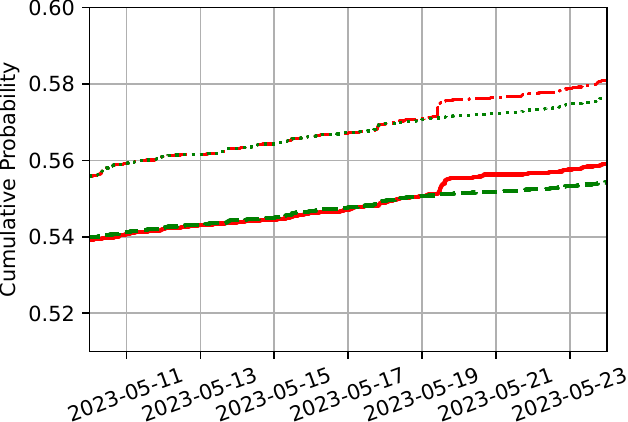}
        \caption{Exclusion of 2023-05-17.}
        \label{fig:net-org-adv:2023-05}
    \end{subfigure}
    \includegraphics[width=0.29\textwidth]{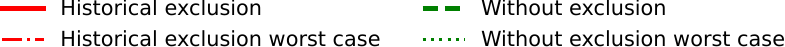}
    \caption{Probability of path compromise with top-3 suspicious \ac{ixp} organizations.
    The gap between the red and green lines shows the advantage the exclusion gave the suspicious
    \acp{ixp} -- roughly a few days of running the network without exclusion.}
    \label{fig:net-org-adv}
\end{figure*}

\begin{figure*}
    \centering
    \begin{subfigure}[b]{0.24\textwidth}
        \centering
        \includegraphics[width=.9\textwidth]{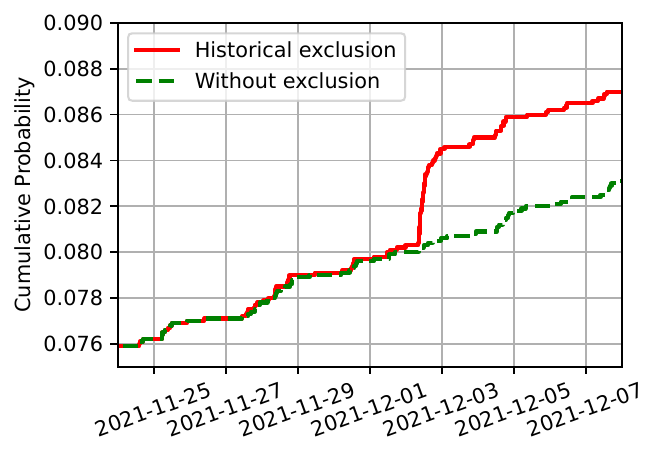}
        \caption{Exclusion of 2021-12-01.}
        \label{fig:results:relay-adv:2021-12}
    \end{subfigure}
    \begin{subfigure}[b]{0.24\textwidth}
        \centering
        \includegraphics[width=.9\textwidth]{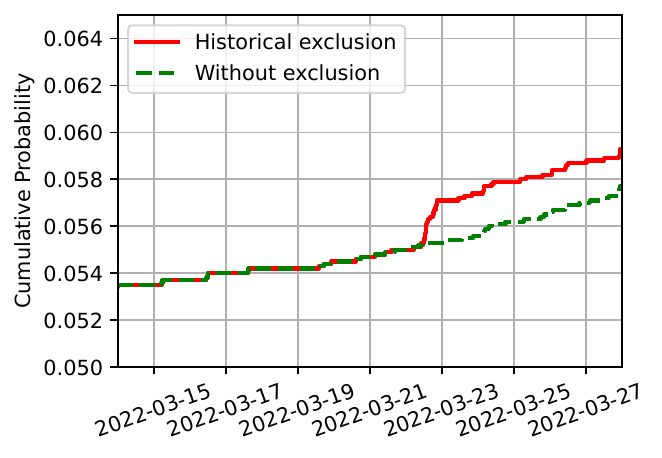}
        \caption{Exclusion of 2022-03-21.}
        \label{fig:results:relay-adv:2022-03}
    \end{subfigure}
    \begin{subfigure}[b]{0.24\textwidth}
        \centering
        \includegraphics[width=.9\textwidth]{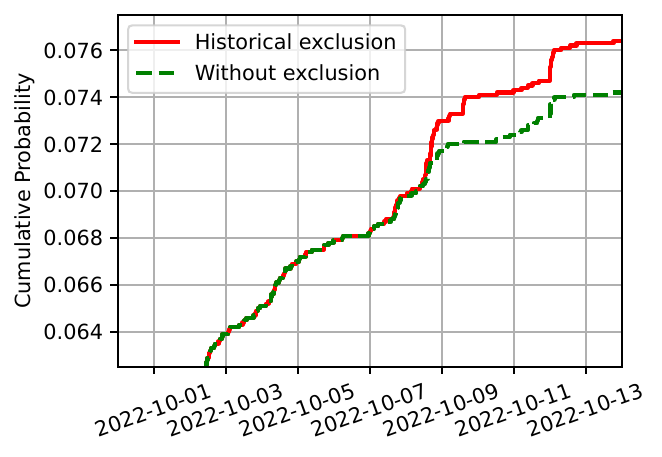}
        \caption{Exclusion of 2022-10-07.}
        \label{fig:results:relay-adv:2022-10}
    \end{subfigure}
    \begin{subfigure}[b]{0.24\textwidth}
        \centering
        \includegraphics[width=.9\textwidth]{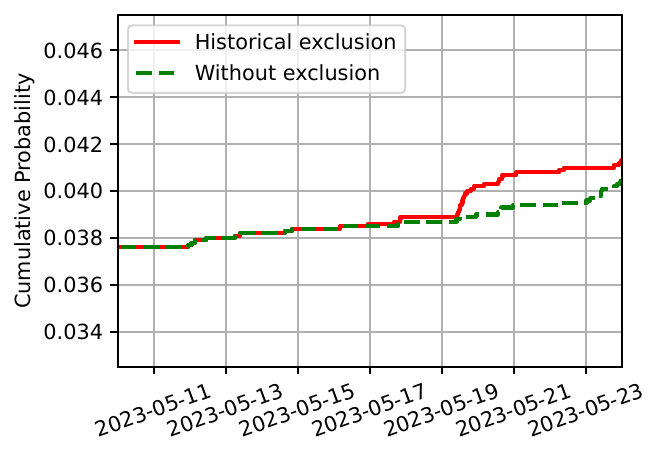}
        \caption{Exclusion of 2023-05-17.}
        \label{fig:results:relay-adv:2023-05}
    \end{subfigure}
    \caption{Probability of path compromise with adversarial relays (10 exits and 20 guards, each with a \num{150000}
    consensus weight).
    The vertical gap between the red and green lines represent the advantage the exclusion gave to the
    adversarial relays: the advantage is equivalent to a few days of running the network without excluding relays.}
    \label{fig:results:relay-adv}
\end{figure*}

While the number of relays listed for exclusion may seem high (ranging from 344 to 644 for the exclusions we
consider)~\cite{tor_eol_policy_2024}, it is noteworthy that the effect of the exclusion is not as alarming.
Although an additional fraction of Tor clients (up to $\approx 1\%$) is suddenly compromised as a result of the
exclusion round, a closer examination of the graph in \cref{fig:results:net-as-adv:2021-12} shows that the
proportion of compromised clients in the historical scenario immediately after the exclusion (on 2021-12-03) is the same as
in the more optimistic scenario where the excluded relays were kept four days later (on 2021-12-07).
This implies that, even without exclusion, the same proportion of compromised clients would have occurred, but four
days later.

Note that the high value for the base compromise rate is due to how the Internet is currently built around high-speed
tier-1 transit \acp{as} such as \as[link=true]{6939} (Hurricane Electric) and \as[link=true]{3356} (Level 3 Parent).
It is possible to reduce this base-rate compromise with path selection logic that leverages the current surplus of
bandwidth within guard and middle positions of circuits to steer paths away from these suspicious \acp{as} while
maintaining a balanced network~\cite{claps}.
Another unexplored approach could be to revisit the guard-flag assignment logic to account for such issues; the
assignment does not even need to be the same for everyone.
It is, however, important to understand that the \textit{how} is not the most difficult problem; addressing this
problem correctly requires tapping into the surplus of bandwidth within path positions, which is equally
useful for solutions to other adversarial threats that the Tor network faces, such as website fingerprinting
potentially addressed by traffic coverage~\cite{prop265, maybenot}, which also exploits bandwidth surplus.
Ultimately, it becomes more a question of \textit{what} should be done rather than \textit{how} to approach these problems
individually.
Possibly, a compromise could be found to tap into the same bandwidth surplus, but \textit{what} compromise is needed
remains an open question.

\subsection{\ac{ixp} adversary}
\label{subsec:ixp-adversary}

Similar to the \ac{as} adversary, the top three organizations that administer multiple \acp{ixp} are considered suspicious.
\Cref{fig:net-org-adv} presents the experimental results.
The same observations made for \ac{as} adversaries apply.

Inferring which \ac{ixp} transfers data between two \acp{as} is not straightforward.
First, previous literature~\cite{augustin_ixps_2009} indicates that detecting an \ac{ixp} between two \acp{as} is
not trivial.
Second, when \acp{as} on the path are known to be connected by multiple \acp{ixp}, there is no simple way to determine which
\ac{ixp} is used, if any.
In fact, even if two \acp{as} are connected by a single \ac{ixp}, those two \acp{as} might also be paired directly.
In that case, the traffic might completely avoid the intermediary \ac{ixp}.
To determine the \ac{ixp} that observes the traffic flowing between a pair of \acp{as}, multiple strategies are possible.

In \cref{fig:net-org-adv}, two selection strategies are depicted.
First, one \ac{ixp} is selected at random when multiple \acp{ixp} are likely to connect two \acp{as}.
Second, all \acp{ixp} that could connect two \acp{as} are selected and presented as the worst case on the graphs in
\cref{fig:net-org-adv}.
That is, when an \ac{ixp} must be selected to connect two \acp{as}, all \acp{ixp} connecting the
two \acp{as} are selected.
As expected, the proportion of compromised paths is slightly higher with this heuristic than when a single
\ac{ixp} is selected at random.
However, the relative difference between the historical exclusion and the simulation without exclusion remains similar.
This indicates that the inference algorithm used to select \acp{ixp} influences the proportion of compromised
paths but does not invalidate our claim.

\subsection{Relay adversary}
\label{subsec:relay-adversary}

For each relay adversary experiment, the adversary controls 10 exit relays and 20 guard relays added to the network.
Each adversary-controlled relay has a consensus weight of \num{150000}, similar to the top relays in the network.
These figures are reasonable assumptions based on previously reported malicious actors~\cite{nusenu_how_2020,nusenu_tracking_2021}.

\Cref{fig:results:relay-adv} shows the cumulative distribution function of clients compromised by the relay
adversary over time.
As with the network adversary, the increase in the number of compromised clients diminishes over time.
This can be explained in the same way as for the network adversary.
However, it should be noted that the consensus weight for the adversary remains the same across all experiments,
while the Tor network grew over time.
Thus, the total fraction of the consensus weight controlled by the adversary decreases over time.

One key difference is worth noting: in \cref{fig:results:relay-adv:2021-12}, the probability of compromise
after the exclusion is not reached as quickly as in \cref{fig:results:net-as-adv:2021-12}.
In \cref{fig:results:net-as-adv:2021-12}, it took four days for the scenario in which the excluded relays are kept in
the network to reach the same level of client compromise as in the historical scenario.
Here, the graph in \cref{fig:results:relay-adv:2021-12} is not wide enough to show when this occurs, but the
raw data indicate that it took nine days (instead of four).
This could be explained by the overall diversity that currently exists in the network against each adversarial model.
The higher the diversity, the lower the adversary's strength.
Our results show that the concentration of paths to a few top \acp{as} is more significant than the concentration of
paths to a few top relays.
This might be addressed by adapting the path selection algorithm, and the literature already explores this
idea~\cite{rochet2017waterfilling}.

\section{Individual relay utility and security contribution}
\label{sec:individual}

In the previous section, we showed that excluding relays gives adversaries (presented in \cref{sec:threat-model}) an
advantage over a situation without exclusion.
We now quantify which relay provides the most security to the network among the relays about to be excluded.
To this end, we propose metrics that capture the utility and security contribution of a specific relay (\cref{rq2}).
This contrasts with security metrics in related work, which usually aim to evaluate the security of the whole
network~\cite{barton2018towards, serjantov2002towards, diaz2002towards, syverson2009m, rochet2017waterfilling}.
However, summing the contribution of each relay would yield an evaluation of the network's security.
Our metrics focus on quantifying relays' contributions rather than evaluating client anonymity directly.
Each metric is designed to capture the security contribution against a specific adversary we consider.
This depends on multiple properties of the relay and the network, and on how those properties are used in the path
selection algorithm.

In the context of exclusions, the goal of these metrics is to indicate which relays are most
important to maintain in the network.

\subsection{Notations}
\label{subsec:notations}

We extend the conventions in notation used by previous literature~\cite{jaggard2017onions, jaggard201520} and the Tor
specification~\cite{tor_spec}.
For a set of relays in the network, we use uppercase calligraphic letters (e.g., the set $\mathcal{G}$ of all relays flagged as
guards in the network).
We use lowercase italicized letters for a particular relay (e.g., a guard $\textit{g}$).

For the remainder of the section, we define the following sets:
\begin{itemize}
    \item $\mathcal{R}$, the set of all relays in the network.
    \item $\mathcal{A}$, the set of all network adversaries.
    \item $\mathcal{G}$, the set of all relays flagged only as guard.
    \item $\mathcal{E}$, the set of all relays flagged only as exit.
    \item $\mathcal{D}$, the set of all relays flagged both as guard and exit.
\end{itemize}

As mentioned in \cref{subsec:torbasics}, for a relay~$\textit{r}$, the bandwidth used in the path selection algorithm is
computed differently depending on the position in the circuit in which we want it to be.
We denote the bandwidth for a relay in the guard position as $\text{bw}_{g}(\textit{r})$.
Likewise, for the exit position, we write $\text{bw}_{e}(\textit{r})$.

We also use the normalized bandwidth of a relay with respect to its position in a circuit.
For instance, let $\textit{r}$ be a relay, and let $\text{bw}_{g}(\textit{r})$ be its bandwidth in the Tor network as a
guard.
Then the normalized bandwidth of that relay is computed with the following formula:
\begin{equation}
    \text{norm}_{\mathcal{G} \cup \mathcal{D}}(\text{bw}_{g}(\textit{r})) =
    \frac{\text{bw}_{g}(\textit{r})}{\sum_{\textit{g} \in \mathcal{G}\cup \mathcal{D}} \text{bw}_{g}(\textit{g})}\text{.}
    \label{eq:norm}
\end{equation}
Note that this normalization is performed over the bandwidth of all the guard flagged relays in the network, not all the
relays.

We also define, for a given guard $\textit{g}$ used in a circuit, the set $\text{reach}_{\mathcal{E}\cup \mathcal{D}}(g)$ corresponding to all
exit flagged relays that can be paired with $\textit{g}$ in the circuit.
Likewise, for a given exit $\textit{e}$ in the circuit, $\text{reach}_{\mathcal{G}\cup \mathcal{D}}(\textit{e})$ is the set of
guard flagged relays that can be paired with $e$ as the guard of a circuit.

\subsection{Relay utility metric}
\label{subsec:relay-utility-metric}

We define relay utility as the number of circuits that can be built using that relay.
This depends on the number of other relays that can be combined with it to complete the circuit.
Instead of considering the raw number of relays, we use the relays' bandwidth.
This captures not only the number of circuits but also how many clients can use those circuits.

This metric does not make strong assumptions about a potential relay adversary.
We assume that relays about to be rejected are not malicious and that the adversary already operates other relays in
the network.
The metric estimates relay utility for clients: useful relays have relatively high bandwidth
and can be paired with a substantial share of exit capacity.
This value also indicates the marginal increase in success for an already operating relay adversary: if a relay is used by a
large number of clients and is excluded, all those clients will have to select new relays for their circuits and might
end up selecting an adversary-operated relay.

When considering a relay $\textit{g}$ as a guard in a circuit, we are interested in the exit relays that can be paired
with it.
For a guard $\textit{g} \in \mathcal{G}$, we evaluate its utility as follows:
\begin{equation}
    \label{eq:rmg}
    \text{RM}(\textit{g}) =
    \text{norm}_{\mathcal{G}\cup \mathcal{D}}(\text{bw}_{g}(\textit{g})) \times \sum\limits_
        {\textit{e} \in \text{reach}_{\mathcal{E}\cup \mathcal{D}}(\textit{g})}
    \text{norm}_{\mathcal{E}\cup \mathcal{D}}(\text{bw}_{e}(\textit{e}))\text{.}
\end{equation}

For an exit $\textit{e} \in \mathcal{E}$, we obtain the following metric using the same reasoning:
\begin{equation}
    \label{eq:rme}
    \text{RM}(\textit{e}) = \text{norm}_{\mathcal{E}\cup \mathcal{D}}(\text{bw}_{e}(\textit{e})) \times \sum\limits_
        {\textit{g} \in \text{reach}_{\mathcal{G}\cup \mathcal{D}}(\textit{e})}
    \text{norm}_{\mathcal{G}\cup \mathcal{D}}(\text{bw}_{g}(\textit{g}))\text{.}
\end{equation}

When a relay $\textit{d} \in \mathcal{D}$ can serve as a guard or an exit relay in a circuit, its total utility is the sum of its utility
as a guard and as an exit.
Note that there is no double counting of resources: $\text{bw}_{g}(\textit{d})$ and $\text{bw}_{e}(\textit{d})$ are
computed by the authorities to distribute the resources of $d$ between the guard and exit positions.
This yields the following metric:
\begin{equation}
    \label{eq:rmd}
    \begin{aligned}
        \text{RM}(\textit{d}) = &\ \ \text{norm}_{\mathcal{G}\cup \mathcal{D}}(\text{bw}_{g}(\textit{d})) \times \sum\limits_
            {\textit{e} \in
        \text{reach}_{\mathcal{E}\cup \mathcal{D}}(\textit{d})}
        \text{norm}_{\mathcal{E}\cup \mathcal{D}}(\text{bw}_{e}(\textit{e}))\\
        + &\ \ \text{norm}_{\mathcal{E}\cup \mathcal{D}}(\text{bw}_{e}(\textit{d})) \times \sum\limits_
            {\textit{g} \in \text{reach}_{\mathcal{G}\cup \mathcal{D}}(\textit{d})}
        \text{norm}_{\mathcal{G}\cup \mathcal{D}}(\text{bw}_{g}(\textit{g}))\text{.}
    \end{aligned}
\end{equation}

Multiple factors govern the value of the relay utility metric:
\begin{inparaenum}[(i)]
    \item the share of bandwidth the relay provides in its position (captured by $\text{bw}_{g}(\textit{r})$ and
    $\text{bw}_{e}(\textit{r})$),
    \item the share of relays that can be used to build a circuit (captured by $\text{reach}_{\mathcal{G}\cup\mathcal{D}}(\textit{r})$ and
    $\text{reach}_{\mathcal{E}\cup \mathcal{D}}(\textit{r})$), and
    \item relay-selection constraints such as families and network addresses (captured by
    $\text{reach}_{\mathcal{G}\cup \mathcal{D}} (\textit{r})$ and $\text{reach}_{\mathcal{E}\cup \mathcal{D}}(\textit{r})$).
\end{inparaenum}
For more details, the reader may refer to \cref{appendix:factors-relay-metric}.

\subsection{Network adversary metric}
\label{subsec:network-adversary-metric}

The second metric we propose measures how unlikely it is that a relay is used to build a compromised circuit.
This is applicable to \ac{as} and \ac{ixp} adversaries.
However, we consider only \ac{as} adversaries in this section.
As mentioned in \cref{subsec:ixp-adversary}, inferring \acp{ixp} along an Internet path depends on the
selected heuristic and can yield very different results.

The idea is to compute the probability that a network adversary $\textit{a}$ appears on both ends of a circuit
given that a relay $\textit{r}$ is used to construct the circuit.
We denote that probability as $\Pr(\textit{a}|\textit{r})$.
This probability is the product of the probability $\Pr(\textit{a}|\textit{g})$ that $\textit{a}$ appears before the
guard $\textit{g}$ and the probability $\Pr(\textit{a}|\textit{e})$ that $\textit{a}$ appears after the exit
$\textit{e}$.
We also use the probability $\Pr(\textit{r})$ that a relay $\textit{r}$ is in a circuit and the probability
$\Pr(\textit{r}|\textit{r}')$ that $\textit{r}$ is in the circuit given that $\textit{r}'$ is in the circuit.

For a relay $\textit{r}$, we then compute the following metric:
\begin{equation}
    \label{eq:nmr}
    \text{NM}(\textit{r}) = \Pr(\textit{r}) \prod_{\textit{a} \in \mathcal{A}}  1 - \Pr(\textit{a}|\textit{r})\text{,}
\end{equation}
capturing, for all adversaries $\textit{a}$, the probability that $\textit{r}$ is in the circuit and does not put
$\textit{a}$ in a position to compromise the client.
In practice, we can replace the probability $\Pr(\textit{r})$ with the bandwidth $\text{bw}(\textit{r})$, as this is
how path selection weights the relays when choosing them randomly.

If we consider a guard $\textit{g} \in \mathcal{G}$, the fully developed metric is as follows:
\begin{equation}
    \text{NM}(\textit{g}) = \text{bw}_{g}(\textit{g}) \prod_{\textit{a} \in \mathcal{A}}  1 - \left[\Pr(\textit{a}|\textit{g})
    \times \sum_{\textit{e} \in \text{reach}_{\mathcal{E}\cup \mathcal{D}}(\textit{g})} \Pr(\textit{a}|\textit{e})
    \Pr(\textit{e}|\textit{g})\right]\text{.}
    \label{eq:nmg}
\end{equation}

Similarly, if we consider an exit $\textit{e}  \in \mathcal{E}$, we obtain:
\begin{equation}
    \text{NM}(\textit{e}) = \text{bw}_{e}(\textit{e}) \prod_{\textit{a} \in \mathcal{A}}  1 - \left[\Pr(\textit{a}|\textit{e})
    \times \sum_{\textit{g} \in \text{reach}_{\mathcal{G}\cup \mathcal{D}}(\textit{e})} \Pr(\textit{a}|\textit{g})
    \Pr(\textit{g}|\textit{e})\right]\text{.}
    \label{eq:nme}
\end{equation}

Finally, if we consider a relay $\textit{d} \in \mathcal{D}$ that can be a guard or an exit relay, its total
contribution is the sum of its contributions as guard and exit.
As with $\text{RM}(\textit{d})$ (\cref{eq:rmd}), there is no double counting of resources, as
$\text{bw}_{g}(\textit{d})$ and $\text{bw}_{e}(\textit{d})$ are computed by the authorities to distribute the
resources of relay $\textit{d}$ between the guard and exit positions.
This yields the following metric:
\begin{equation}
    \begin{aligned}
        \text{NM}(\textit{d}) = &\ \ \text{bw}_{g}(\textit{d}) \prod_{\textit{a} \in \mathcal{A}}  1 -
        \left[\Pr(\textit{a}|\textit{d})\times \sum_{\textit{e} \in \text{reach}_{\mathcal{E}\cup \mathcal{D}}(\textit{d})}
        \Pr(\textit{a}|\textit{e})\Pr(\textit{e}|\textit{d})\right]\\
        + &\ \ \text{bw}_{e}(\textit{d}) \prod_{\textit{a} \in \mathcal{A}}  1 - \left[\Pr(\textit{a}|\textit{d})\times \sum_
            {\textit{g} \in \text{reach}_{\mathcal{G}\cup \mathcal{D}}(\textit{d})} \Pr(\textit{a}|\textit{g})\Pr(\textit{g}|
        \textit{d})\right]
        \text{.}
    \end{aligned}
    \label{eq:nmd}
\end{equation}

The different probabilities can be computed at a given time based on public data from the consensus, knowledge of
Tor's path selection algorithm, and the \ac{as} path inference used in \cref{sec:anonymity-impact} once the
distribution of paths is known.

\begin{figure}
    \centering
    \begin{subfigure}[t]{0.48\textwidth}
        \centering
        \includegraphics[scale=.4]{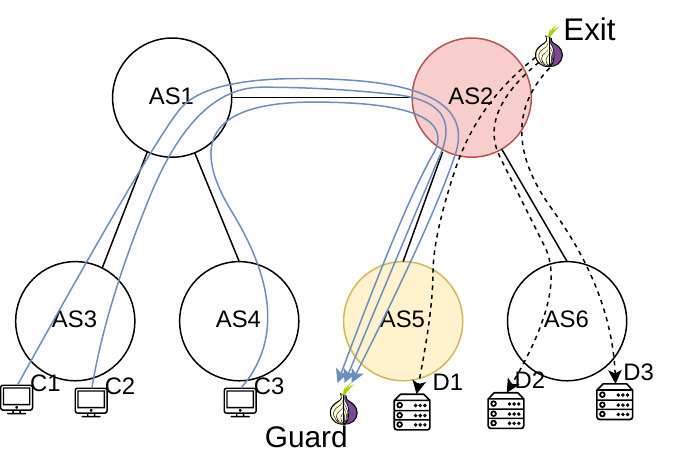}
        \caption{Different relay positions. \as{2} can see all flows, and \as{5} can see one flow.}
        \label{fig:net_metric_example_gard_pos}
    \end{subfigure}

    \begin{subfigure}[t]{0.48\textwidth}
        \centering
        \includegraphics[scale=.4]{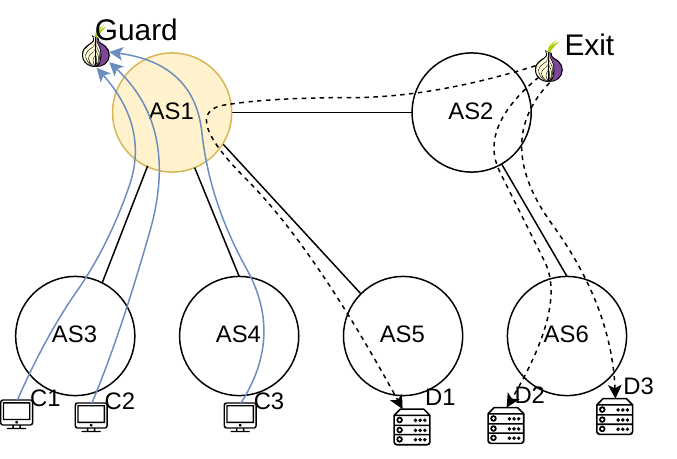}
        \caption{Changing the Internet topology. \as{1} can correlate one flow.}
        \label{fig:net_metric_example_topo}
    \end{subfigure}

    \begin{subfigure}[t]{0.48\textwidth}
        \centering
        \includegraphics[scale=.4]{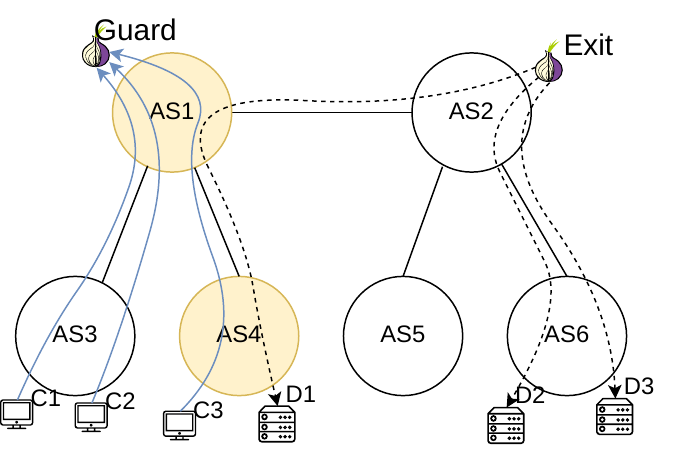}
        \caption{Changing the destination position. \as{1} and \as{4} can both correlate the same flow.}
        \label{fig:net_metric_example_dest_pos}
    \end{subfigure}
    \caption{Examples of factors influencing the network adversary metric. Colored \acp{as} can correlate the traffic.}
    \label{fig:net_metric_example}
\end{figure}

Multiple network factors can influence the result of our network adversary metric.
\Cref{fig:net_metric_example} illustrates variants of a small example network involving a guard and an exit
alongside some clients and destinations those clients intend to visit.
In a perfect situation for the guard according to our metric, all possible client paths to reach the guard never
traverse an \ac{as} that can also intercept traffic flowing from an exit to a potential destination.
The solid blue arrows indicate all the paths used by client data to reach the guard.
Similarly, the dashed arrows indicate all the paths used by client data to reach their destination from the exit relay.
All those paths traverse the topology of \acp{as} used in this example.
For each variation, we show how the guard's metric changes due to the modification in the network.
To make the example clearer, we chose not to represent more than one exit that can be paired with the guard.
In practice, the metric would be computed by taking into account all exits in the network that can be paired with the
guard to form a circuit.

\Cref{fig:net_metric_example_gard_pos} illustrates how the guard's position in the network can influence its metric.
In that situation, the guard is placed in \as{5}.
Now two \acp{as} have a nonzero probability of being in a position to correlate traffic.
\as{5} can now see, with probability $1/3$, the traffic on both ends.
Even worse, \as{2} has probability $1$ of being in a position to correlate the traffic.
This gives the guard a network adversary metric of $\text{NM}(\textit{g}) = 0$.

\Cref{fig:net_metric_example_topo} illustrates how the Internet topology can also influence the guard's network adversary metric.
In this example, \as{5} is connected to \as{1} instead of \as{2}.
This changes the path from the exit relay to destination \texttt{D1}.
Thus, \as{1} now has probability $1/3$ of intercepting both ends of a circuit.
For that reason, the guard's metric is now $\text{NM}(\textit{g}) = \text{bw}_{g}(\textit{g}) \times 2/3$.

Finally, in \Cref{fig:net_metric_example_dest_pos}, we illustrate the influence of the destination's position in the
network.
Here, destination \texttt{D1} has moved from \as{5} to \as{4}.
In that case, \as{1} and \as{4} are both able to perform traffic correlation.
For \as{1}, the probability that both ends of a circuit traverse it equals $1/3$, and for \as{4} that probability
equals $1/3 \times 1/3 = 1/9$.
This yields a final metric of $\text{NM}(\textit{g}) = \text{bw}_{g}(\textit{g}) \times 16/27$ for the guard.
Symmetrically, it is clear that the clients' positions also contribute to that guard's metric.

\subsection{Evaluating excluded relays' contribution}
\label{subsec:evaluating-excluded-relays-contribution}

\begin{figure}
    \centering
    \begin{subfigure}[b]{0.23\textwidth}
        \centering
        \includegraphics[width=\linewidth]{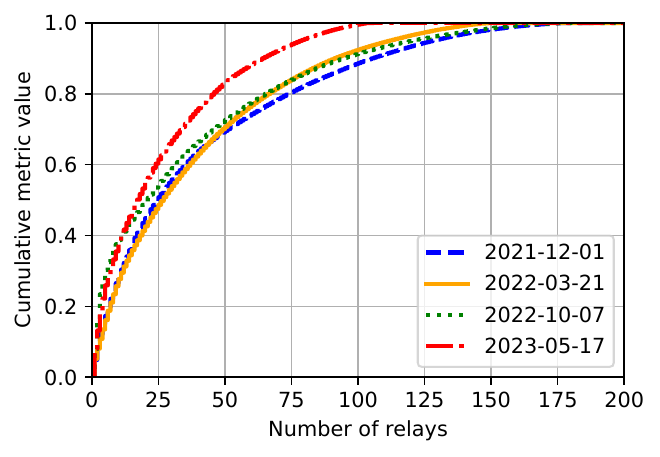}
        \caption{Relay metric.}
        \label{fig:relay_metric}
    \end{subfigure}
    \begin{subfigure}[b]{0.23\textwidth}
        \centering
        \includegraphics[width=\linewidth]{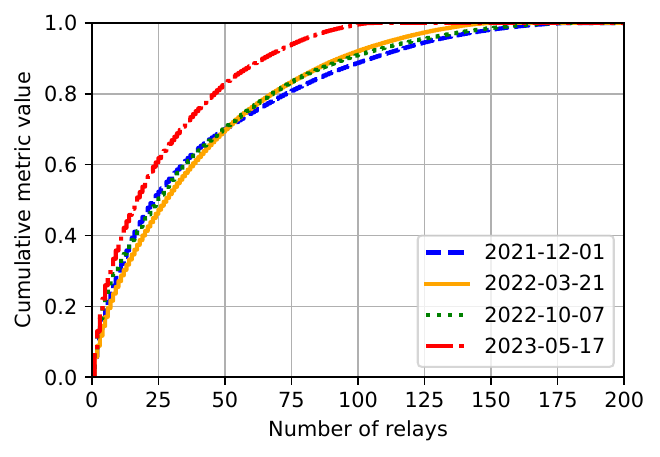}
        \caption{Network metric.}
        \label{fig:network_metric}
    \end{subfigure}
    \caption
    {Cumulative value for the relay metric (\cref{eq:rmg,eq:rme,eq:rmd}) and network metric
        (\cref{eq:nmg,eq:nme,eq:nmd}). For each date, only the relays set for
    exclusion are considered.}
    \label{fig:metrics-cdf}
\end{figure}

We compute both relay and network metrics for the relays set for the four exclusion rounds: 2021-12-01, 2022-03-21,
2022-10-07, and 2023-05-17 to illustrate how they can help identify important relays to keep in the network.

\subsubsection{Relay adversary metric.}
\label{subsubsec:relay-adversary-metric}

\Cref{fig:relay_metric} shows the cumulative distribution functions of the relay adversary metric for the four
exclusion rounds mentioned earlier.
Each curve in the figure represents the proportion of all excluded relays' contributions to resisting relay adversaries.
For example, the curve for the exclusion round in May 2023 indicates that 25 relays contributed 60\% of the total
resilience to relay adversaries provided by all excluded relays.

\subsubsection{Network adversary metric.}
\label{subsubsec:network-adversary-metric}
\Cref{fig:network_metric} shows the cumulative distribution for the metric over all the relays listed for exclusion.
It is worth noting that, to cover 50\% of the cumulative metric, it is generally sufficient to ensure that the 25
\say{largest} relays according to this metric are updated before the exclusion.
Contacting all the relay operators is not as critical, as a few relays account for a large share of the cumulative
value of the metric.

The first three exclusion rounds represented on the graph have similar curves.
However, the curve for the exclusion of May 2023 is shifted to the left: a smaller set of relays accounts for the
same share of the cumulative metric that will be excluded.
The exclusion of May 2023 is slightly smaller than that of March 2022 (344 relays and 354 relays, respectively).
As the metric has a value of zero for all middle relays, the leftward shift of the May 2023 curve
is caused by a smaller number of non-middle-only relays being excluded (106 for May 2023 versus 182 for March 2022).

\subsubsection{Ranking relays.}

We use the metrics to rank relays about to be excluded from the network.
To provide insight into how the metrics can identify important relays, we list the 15 relays with the highest scores
for each metric in~\cref{tab:relay_top15,tab:net_top15}.
We depict only the exclusion of October 2022.

The tables in \Cref{appendix:ranking} show that relays with higher bandwidth generally appear at
the top of the ranking: bandwidth is part of the computation for both metrics and is directly linked to the number
of clients that the relay can handle.
However, both metrics provide additional nuance: sorting by the metric does not simply sort by bandwidth.
The position the relay can take in a circuit is also taken into account: most exit relays are pushed to the top of
the ranking, especially in~\cref{tab:relay_top15}.

It is interesting to note that the relay \texttt{xeracition} is in the top-3 in both tables with only 63\% of the
bandwidth of the first relay (\texttt{slauthtertodeport1}).
For the utility metric (\cref{eq:rme}), this means its normalized bandwidth as an exit is much lower than
\texttt{slauthtertodeport1}, but it can connect to more guard capacity, as the final metric value is close to the
metric of \texttt{slauthtertodeport1}.
Likewise, for the network metric (\cref{eq:nme}), \texttt{xeracition} is better at avoiding network-level
adversaries than \texttt{slauthtertodeport1}: it has nearly the same metric value with significantly lower bandwidth.
Another interesting relay is \texttt{Password}: this is the only relay with less than 30~MiB/s of bandwidth
to be ranked in \cref{tab:relay_top15}.
These results show that unusual locations of some relays have a significant positive impact on the network's provided
security against our modeled adversaries.
Potentially, such a metric and ranking could be part of incentives to foster diversity.

Looking at the most popular \acp{as} appearing in \cref{tab:relay_top15,tab:net_top15}, one of them stands out.
\as[link=true]{16276} (OVH, a popular French hosting provider) appears five times in \cref{tab:relay_top15} and six
times in \cref{tab:net_top15}.
The popularity of OVH across relay operators does not help the relays achieve large metric values.
The Tor path selection algorithm prevents any two relays hosted in the same \texttt{/16} subnet (or \texttt{/32} for
IPv6) from being used in the same circuit.
This restriction is captured in both metrics and lowers the score despite their high bandwidth.
Having a large share of relays hosted in a single \ac{as} especially lowers the score for the network adversary metric,
as this increases the probability of this \ac{as} being able to observe both ends of the circuit.
In the particular case of OVH, the provider is also popular for web hosting, so destinations to which Tor clients
connect are also likely to be observable by this \ac{as}.
For those reasons, it is reasonable that relays hosted at OVH are ranked lower in \cref{tab:relay_top15,tab:net_top15}
than other relays with similar bandwidth.

From the tables, we see that, in general, relays with a high metric value for one metric also have a high value for
the other metric: 11 relays appear in both tables.
However, four relays appear in only one table, indicating that the metrics effectively capture different aspects.

Finally, the human factor in relay operation is also visible: multiple relays from the same family were set for
exclusion and appear in the top 15 both~\cref{tab:relay_top15,tab:net_top15}.
Relays from the same family are operated by the same person or the same organization.
Similarly, multiple relays were quite old (3+ years), indicating that the operator already has some experience in
operating Tor relays but may suffer from the \say{up-to-date within the official repository but rejected from the
Tor network} issue, as discussed in \cref{sec:social-aspects}.

\subsection{Discussion}
\label{subsec:discussion}

While the metrics presented in \cref{subsec:relay-utility-metric,subsec:network-adversary-metric} are based on the
consensus weight and other factors (such as network location and flags), we observed that the consensus weight is the
most significant component of the metric on average.
That is, a relay's consensus weight is a good estimate for our suggested metrics in the average case, given the current
Internet structure and relay distribution relative to the adversary's position.
Either may change in the future.
This observation also depends on how we reason about adversaries: our analysis considers that any \ac{as} on the
path might potentially be malicious.
However, if the adversarial \acp{as} or victims are known~\cite{jaggard2017onions}, the metrics may depend less on
consensus weight and more on network topology and the location of those adversarial \acp{as} and victims.
\section{Limitations}
\label{sec:dns-resolution}

To create the list of popular destinations (\cref{sec:anonymity-impact}), we used the list produced by
Tranco~\cite{le_pochat_tranco_2019}.
The list provides only domain names, whereas Tor circuits are built to reach a specific IP address.
To this end, we resolved the domains found in the Tranco lists using our DNS server.
Because of the prevalent use of \acp{cdn}, the IP addresses obtained when resolving the domain names may depend on
our location.
This limitation is common to most papers using such a
list~\cite{jueckstock_towards_2021,rimmer2022trace,niaki_iclab_2020} and to widely used systems for Tor simulation,
such as Shadow~\cite{jansen2022co, shadow-ndss12}, which also resolve IPs from a small set of vantage points.

In addition to the \ac{cdn} issue, all DNS resolutions were performed in 2024 (at the time of the experiments).
Thus, domains that were popular in 2019 were resolved in 2024 and may no longer have the same IP addresses or may simply
no longer exist.
However, the methodology we introduce, if used by the Tor Project to analyze imminent relay rejection, would not
face this limitation.

Considering this, the AS-paths constructed between the exit nodes and the destinations in our experiments may not be
the same for all destinations when using another vantage point in space or time.
The potential impact of this is not clear, as already discussed by other simulation tools for Tor~\cite{parsealexa_py}.

Such limitations could be mitigated using a crowdsourced DNS measurement approach with data retention.
Clients distributed across the world could periodically resolve popular domains and share the responses they obtained,
as well as some network location data.
Initiatives like OONI~\cite{ooni} or RIPE Atlas~\cite{ripe_atlas} could be leveraged to implement such crowdsourced data
collection and to provide historical results similar to the Tor Project's CollecTor archive~\cite{collector} for Tor data.

While OONI and RIPE Atlas offer DNS resolution capabilities, their application to large-scale DNS measurement presents challenges.
OONI's domain list is relatively static, with no straightforward mechanism for frequent updates.
RIPE Atlas, in contrast, requires users to earn credits (earned by running probes) which constrains query volume
based on both probe availability and credit accumulation.
Moreover, RIPE Atlas imposes rate limits on DNS resolution requests per probe within defined timeframes.

\section{Ethical considerations}
\label{sec:ethics}
Interviews followed university guidelines compliant with GDPR, with participants giving informed consent.
As stated in the consent form, audio recordings of the interviews have been deleted since the publication of this paper
(September 2026), and transcriptions of the interviews will be retained for a maximum of two years after publication.
Only the research team accessed the data, and participants could skip questions or withdraw, with their data destroyed if they did so.
This was explained beforehand in the interview guide (\cref{sec:interview_guide}).
Results were reported with aggregated numbers and anonymized comments.

All experiments were conducted in controlled simulators, interacting only with the live Internet for DNS resolution using institutional servers.

\section{Conclusion}
\label{sec:conclusion}

Since 2019, the Tor Project has rejected outdated relays after each new major release.
In some cases, the number and bandwidth of rejected relays seemed concerning despite Network Health Team mitigation efforts.
We study this policy's impact (\cref{rq1}) from relay operators' perspective (\cref{contrib1}, \cref{sec:social-aspects})
and via network simulations (\cref{contrib2}, \cref{sec:anonymity-impact}).

Interviews with 26 Tor relay operators revealed that manual updates are common, even among large operators.
Some used default package repositories for Debian and Ubuntu, despite Tor Project recommendations
against this practice~\cite{torproject_installing}, leading to exclusion  from the network even when their
distributions were still supported.
While operators acknowledged the security benefits of the \ac{eol} policy, some noted concerns about increased
operator burden and reduced network diversity.

The operators we interviewed are probably among the most committed: the vast majority of them were never excluded from
the network and would understand if one of their relays were rejected because it was too out of date.
However, when analyzing the data related to the April 2024 exclusion, we note that 85\%
of excluded relays did not return with an updated version within a month.

Based on the simulations, we showed that past rejections had a limited immediate security impact on Tor clients.
However, there is a potentially more concerning and persistent community impact on operators, since we observe that
most excluded relays did not return.
We also observe that although the rejections strengthened the adversary's success, they merely accelerated it by a few days.
Our results provide evidence that the churn of existing relays is an important factor in reducing the anonymity of Tor
clients.

Relay rejections may conflict with existing \say{up-to-date} Tor versions within \ac{os}
distributions, which may cause new operators to install a Tor version that is immediately rejected.
There is no simple technical solution to a problem inherently caused by policy misalignment among independent software
projects.
An ambitious direction may be to redesign the lifecycle of the Tor software to make it independent of the systems on
which it runs.
A proof-of-concept~\cite{rochet2022towards} supports controlling the software lifecycle by hot-swapping updates
in memory via JITed bytecode, but introduces its own downsides and open questions.

We also propose an approach to measure the security contribution provided by individual Tor relays (\cref{rq2})
using security metrics (\cref{contrib3}, \cref{sec:individual}).
We design these metrics based on two commonly considered adversaries against Tor: a network adversary controlling part
of the network infrastructure and a relay adversary adding malicious relay capacity to Tor.
We show how the metrics can be used to rank relays about to be rejected from the network and provide insight into the
most important relays in the network.

Using insights from qualitative analysis, simulations, and relay ranking, we present three key
recommendations addressed to the Tor Project developers to facilitate \ac{eol} policy implementation.
\begin{enumerate}
    \item \textbf{Consider not distributing Tor via official distribution package repositories known to fall
    out of date regularly} (e.g., Debian, Ubuntu, and possibly Alpine, based on \cref{fig:tor-pkg-version}).
    This approach would prevent accidental installation of outdated Tor versions and require new operators to read Tor
    documentation, which advises against using the default package manager for Debian and Ubuntu~\cite{torproject_installing}.

    \item \textbf{Keep the \ac{eol} policy clear for operators.}
    \textit{
        When will my relay be excluded if I do not update?
        What happens next?
        How can I rejoin the network?
        How frequently am I supposed to update?
    }
    The policy should explicitly answer these questions for operators.
    Currently, the information is dispersed across multiple documents~\cite{tor_eol_policy_2024,tor_project_expectations_2025}
    and does not address all questions from the operators' perspective.

    \item \textbf{Provide clear, early, and predictable notifications about upcoming rejections to affected operators.}
    Multiple channels (mailing lists, social networks, logs, monitoring tools) are already used, but there is
    documentation describing where operators can subscribe to only receive information about relay updates.
    The mailing lists announcing relay updates also announce other topics (e.g.\ Tor Browser updates) which are not
    relevant to relay operators.
    We suggest using a specific mailing list for those announcements, a dedicated hashtag on social networks, and an
    RSS feed.
\end{enumerate}

\begin{acks}
The authors would like to thank Julien \textsc{Albert} for his input to the qualitative analysis method and the interview process.

Partially funded by the CyberExcellence project of the Public Service of Wallonia (SPW Recherche), convention No.~2110186.
\end{acks}
\FloatBarrier

\bibliographystyle{ACM-Reference-Format}
\balance
\bibliography{references}


\appendix
\crefalias{section}{appendix}
\section{Open science}
\label{sec:anonymized-source}

The tools used to produce the results presented in this paper are available here online.
Note that those tools were not part of the peer-review process. \\
\href{https://github.com/dont-kill-my-relay/relay-rejection}{\texttt{https://github.com/dont-kill-my-relay/relay-rejection}}

\section{Generative AI usage}\label{sec:genai}
We used generative AI to bootstrap minor parts of the code, proofread the text, and rephrase parts of it.
Other than that, generative AI \textbf{was not} used to produce the research output or to write the paper.

\section{Interview process}
\label{app:interview-process}

\subsection{Participant recruitment}

We recruited participants via two channels: the Tor relay operators mailing list~\cite{dejaeghere_tor-relays_2026} 
and paper signs displayed at FOSDEM~\cite{as_396507_emeraldoniondisobeynet_looking_2026}.
The consent form was attached to the email on the mailing list so participants could read it and decided for
themselves to take part in interview.
For participants contacting us during FOSDEM, we shared the consent form using the same channel the participant used
to contact us.
We answered participant questions before starting the interview.
Questions raised on the mailing list were addressed by replying the to the mailing list and can be found in the
online archive of that list~\cite{dejaeghere_tor-relays_2026}.

\subsection{Interview guide}
\label{sec:interview_guide}

Below is the information provided and the questions posed to participants during the interviews.

\paragraph{Introduction}

\begin{enumerate}
    \item Questions are open-ended and have no right or wrong answers.
    \item You may skip any question.
    Simply reply \say{I don't know} or \say{I would rather not answer}, and we will move on.
    \item Take the time you need to reply and gather your thoughts.
    \item You may stop the interview at any time.
    Simply tell me, and I will not save or use your answers.
    \item The interview is structured in three parts:
    \begin{enumerate}
        \item The first concerns your background as a Tor relay operator.
        \item The second concerns your installation and update process in relation to Tor.
        \item The last concerns the \ac{eol} policy applied by the Tor Project since 2019.
    \end{enumerate}
\end{enumerate}

\paragraph{Tor relay operator background}

\begin{enumerate}
    \item How many relays do you operate?
    \item How much total bandwidth do they provide?
    You may specify the total advertised or the total actually used, whichever is easier for you to find or remember.
    \item Since when have you operated these relays?
    \item How are the relays hosted?
\end{enumerate}

\paragraph{Installation and update process}

\begin{enumerate}
    \item How did you install Tor on your relays?
    \item How do you keep your relays up to date with the latest Tor version?
    \item
    In your view, what happens between the moment the Tor Project releases a new version and the moment the update
    is applied to your relay?
    Specifically, who is involved?
    What are the steps?
    \item What are the main benefits and disadvantages of this approach?
\end{enumerate}

\paragraph{\Ac{eol} policy and exclusion}

At this point, and to provide all participants with the same context, the following text is read aloud verbatim:
\itwquote{Since 2019, Tor has been rejecting relays running outdated versions from the network. Currently, only non-\ac{eol}
version series are accepted. When a version series becomes \ac{eol}, it is rejected from the network soon after.
This has resulted in between one and three version series being accepted in the network since 2019.}

During the first interviews, the wording was:
\itwquote{Since 2019, Tor started to reject relays running outdated versions from the network.
    Currently, only the three most recent version series are accepted.
    When a new version series is released, the oldest accepted version series is rejected from the network.}
A participant pointed out that this did not exactly reflect reality; hence the change.

\begin{enumerate}
    \item Were you aware of this policy before starting the interview?
    \item What is your opinion of this policy?
    \item Has one of your relays ever been excluded from the network because it was too out dated?
    \begin{enumerate}
        \item If so, how did you feel, and how did you react?
        \item If not, how would you feel and react if one of your relays were rejected from the Tor network?
    \end{enumerate}
\end{enumerate}
\section{Factors influencing the relay metric}
\label{appendix:factors-relay-metric}

\begin{figure}[b]
    \centering
    \begin{subfigure}[t]{0.48\textwidth}
        \centering
        \includegraphics[scale=.65]{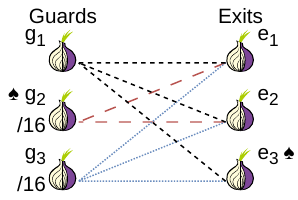}
        \caption{Initial network setup to evaluate the relay utility metric.}
        \label{fig:relay_metric_initial}
    \end{subfigure}

    \begin{subfigure}[t]{0.48\textwidth}
        \centering
        \includegraphics[scale=.65]{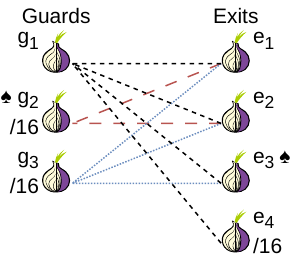}
        \caption{Adding an exit relay $e_4$ decreases the utility of guards $g_2$ and $g_3$.}
        \label{fig:relay_metric_exit}
    \end{subfigure}

    \begin{subfigure}[t]{0.48\textwidth}
        \centering
        \includegraphics[scale=.65]{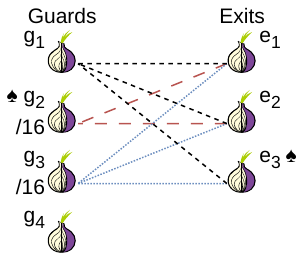}
        \caption{Adding a new guard relay $g_4$ decreases the utility of all guards.}
        \label{fig:relay_metric_guard}
    \end{subfigure}

    \begin{center}
    \begin{tabular}{cl}
        {\begingroup\normalfont
  \raisebox{-.1\ht\strutbox}{\includegraphics[height=\ht\strutbox]{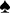}}%
  \endgroup} & Relays share the same family \\
        \texttt{/16} & Relays share the same subnet \\
    \end{tabular}
    \end{center}
    \caption{Examples of factors influencing the relay adversary metric.}
    \label{fig:relay_metric_example}
\end{figure}

Multiple properties of the relays and the network can influence the value of our relay utility metric, presented in
\cref{subsec:relay-utility-metric}.
\Cref{fig:relay_metric_example} highlights influencing factors with a small network setup.
Unless otherwise specified, we assume that all relays have the same bandwidth $b$ for this example.
Also, the relay metric depends on normalized weights but is not normalized itself.
As such, the sum of the metrics, even for the same set of relays, does not necessarily equal one.

In the setup depicted in \cref{fig:relay_metric_initial}, $g_1$ holds one third of the total guard bandwidth
and can connect to all exits.
Therefore, the following computation holds for its utility metric: $\text{RM}(g_1) = 1/3 \times 1$.
The same applies to $g_3$.
For $g_2$, it can connect to only two thirds of the exits.
As expected, its utility is lower because the number of possible circuits is lower: $\text{RM}(g_2) = 1/3 \times 2/3 = 2/9$.

Intuitively, guards $g_1$ and $g_3$ are more useful to clients than $g_2$: they provide access to one more exit
relay.
For the same reason, an adversary already operating in the network would prefer to see $g_1$ or $g_3$ excluded rather
than $g_2$, as more clients would have to re-sample relays to build new circuits.
This is especially true for guards, as they are long-lived: a client attempts to keep using the same guard for an extended
period to avoid being compromised~\cite{dingledine_one_2014}.

If $g_2$ has double the bandwidth (it now has $2b$) and other relays remain unchanged, we have:
$\text{RM}(g_1) = 1/4 \times 1$, $\text{RM}(g_3) = 1/4 \times 1$, and $\text{RM}(g_2) = 1/2 \times 2/3 = 1/3$.
Relay $g_2$ now has a higher utility metric because its bandwidth is twice that of all the other relays,
although it can connect to only two thirds of the exit capacity.

\Cref{fig:relay_metric_exit} (with all relays having the same bandwidth $b$) shows the importance of normalization
in our metric.
Adding one exit relay $e_4$ that can be reached only by $g_1$ will not affect the utility metric of $g_1$: this
guard will still be able to connect to all exits and holds one third of the guard bandwidth (hence its metric
remains $\text{RM}(g_1) = 1/3 \times 1$, as in~\cref{fig:relay_metric_initial}).
However, the metric for both $g_2$ and $g_3$ is now lower: neither can connect to $e_4$, as they are hosted in
the same subnet.
Their metrics are now, respectively, $\text{RM}(g_2) = 1/3 \times 1/2 = 1/6$ and $\text{RM}(g_3) = 1/3 \times 3/4 = 1/4$.

Finally, \cref{fig:relay_metric_guard} shows how adding a guard lowers the utility of all other guards
compared to the setup in \cref{fig:relay_metric_initial}.
Assuming that the new guard also has a bandwidth of $b$, $g_1$ and $g_3$ have a metric of $1/4 \times 1$, and $g_2$
has a metric of $1/4 \times 2/3 = 1/6$.
As there are now more guards in the network, the utility of each guard is relatively lower.
Similarly, the marginal success increase of an already operating adversary will be lower, as clients will be
distributed over more guards.

\section{Metric ranking}
\label{appendix:ranking}

\begin{table*}
    \begin{subtable}{\textwidth}
        \centering
    \begin{threeparttable}
    \begin{tabular}{rlrllS[table-format=-1.2,table-number-alignment = left]lrl}
        \toprule
        \textbf{\#} & \textbf{Relay (family)} & \textbf{$RM$}\hyperref[note:table-1]{~\tnote{a}} &
        \textbf{Location (accuracy)} & \textbf{\ac{as}} &
        \textbf{Bandwidth} & \textbf{First seen} & \textbf{Uptime} & \textbf{Flags}
        \\
        \midrule
        2 & \textcolor{blue}{\textbf{slauthtertodeport1}} & \num{6327184} & LT (200 km) & \as[link=true]{209588} & 75.31~MiB/s & 2021-12-08 & 93 days &
        {\begingroup\normalfont
  \raisebox{-.1\ht\strutbox}{\includegraphics[height=\ht\strutbox]{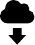}}%
  \endgroup} {\begingroup\normalfont
  \raisebox{-.1\ht\strutbox}{\includegraphics[height=\ht\strutbox]{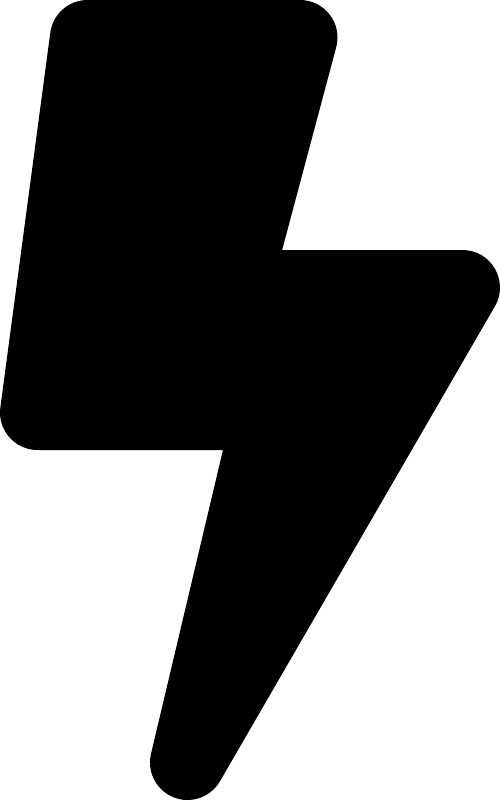}}%
  \endgroup} {\begingroup\normalfont
  \raisebox{-.1\ht\strutbox}{\includegraphics[height=\ht\strutbox]{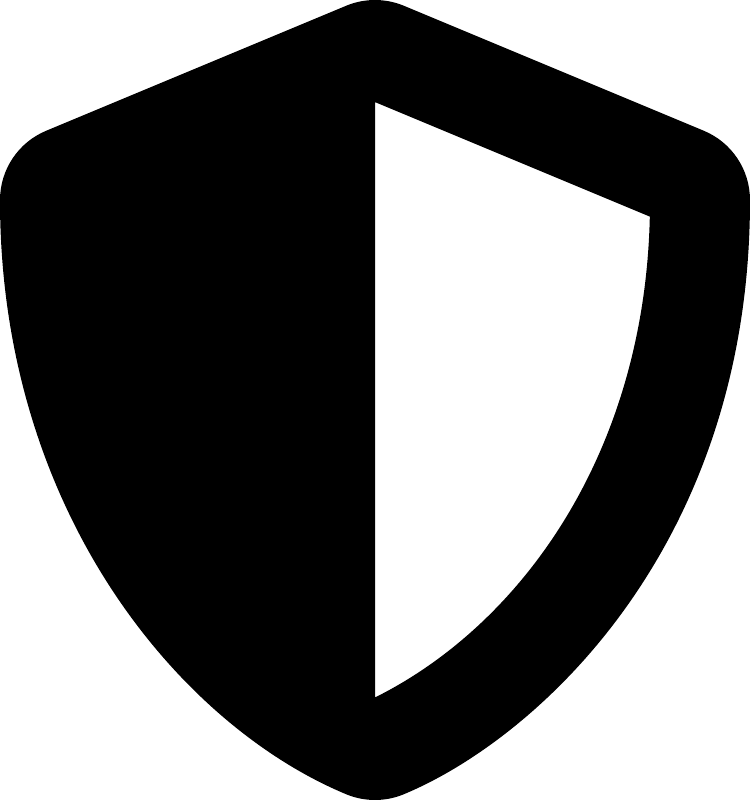}}%
  \endgroup}
        {\begingroup\normalfont
  \raisebox{-.1\ht\strutbox}{\includegraphics[height=\ht\strutbox]{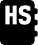}}%
  \endgroup} {\begingroup\normalfont
  \raisebox{-.1\ht\strutbox}{\includegraphics[height=\ht\strutbox]{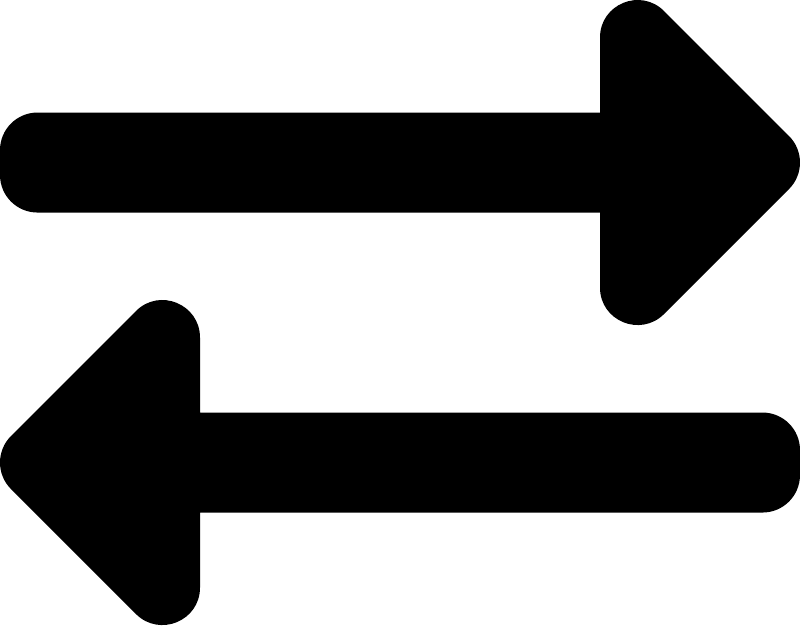}}%
  \endgroup} {\begingroup\normalfont
  \raisebox{-.1\ht\strutbox}{\includegraphics[height=\ht\strutbox]{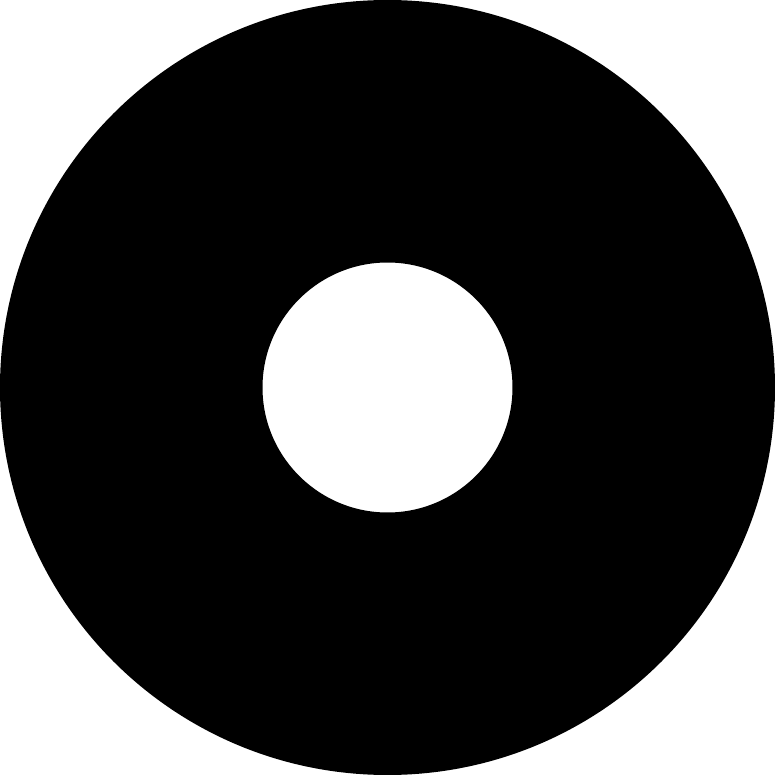}}%
  \endgroup}
        {\begingroup\normalfont
  \raisebox{-.1\ht\strutbox}{\includegraphics[height=\ht\strutbox]{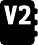}}%
  \endgroup} {\begingroup\normalfont
  \raisebox{-.1\ht\strutbox}{\includegraphics[height=\ht\strutbox]{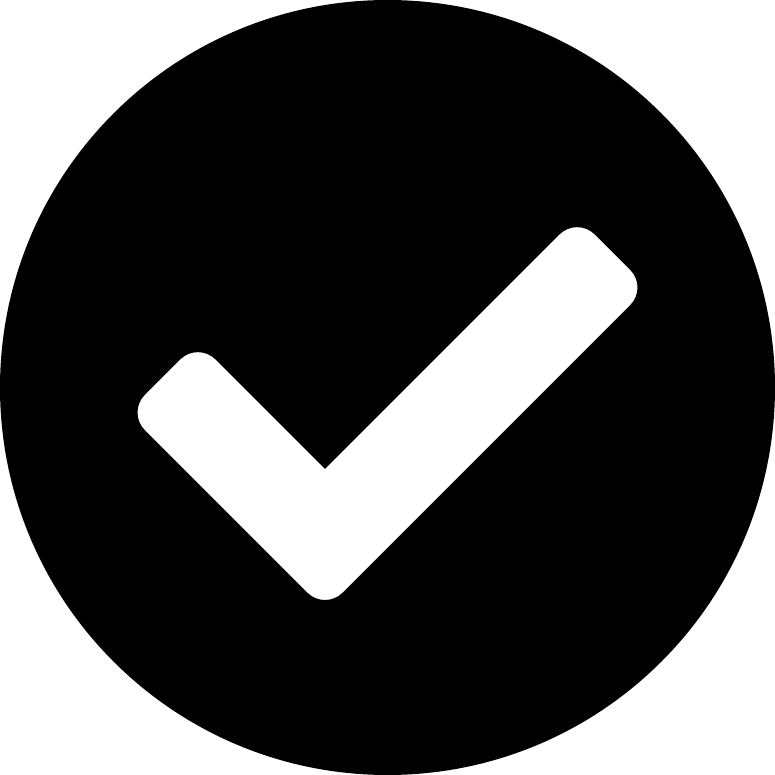}}%
  \endgroup} \\

        3 & \textcolor{blue}{\textbf{xeracition}} & \num{6303121} & DE (200 km) & \as[link=true]{51862} & 47.11~MiB/s & 2022-09-08 & 2 days &
        {\begingroup\normalfont
  \raisebox{-.1\ht\strutbox}{\includegraphics[height=\ht\strutbox]{figures/flags/exit_svg-tex}}%
  \endgroup} {\begingroup\normalfont
  \raisebox{-.1\ht\strutbox}{\includegraphics[height=\ht\strutbox]{figures/flags/fast_svg-tex}}%
  \endgroup} {\begingroup\normalfont
  \raisebox{-.1\ht\strutbox}{\includegraphics[height=\ht\strutbox]{figures/flags/guard_svg-tex}}%
  \endgroup}
        {\begingroup\normalfont
  \raisebox{-.1\ht\strutbox}{\includegraphics[height=\ht\strutbox]{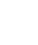}}%
  \endgroup} {\begingroup\normalfont
  \raisebox{-.1\ht\strutbox}{\includegraphics[height=\ht\strutbox]{figures/flags/running_svg-tex}}%
  \endgroup} {\begingroup\normalfont
  \raisebox{-.1\ht\strutbox}{\includegraphics[height=\ht\strutbox]{figures/flags/stable_svg-tex}}%
  \endgroup}
        {\begingroup\normalfont
  \raisebox{-.1\ht\strutbox}{\includegraphics[height=\ht\strutbox]{figures/flags/v2dir_svg-tex}}%
  \endgroup} {\begingroup\normalfont
  \raisebox{-.1\ht\strutbox}{\includegraphics[height=\ht\strutbox]{figures/flags/valid_svg-tex}}%
  \endgroup} \\

        4 & \textcolor{blue}{\textbf{NLfreedom1}} & \num{5976839} & Dronten, NL (20 km) & \as[link=true]{60404} & 68.20~MiB/s & 2022-02-27 & 222 days &
        {\begingroup\normalfont
  \raisebox{-.1\ht\strutbox}{\includegraphics[height=\ht\strutbox]{figures/flags/exit_svg-tex}}%
  \endgroup} {\begingroup\normalfont
  \raisebox{-.1\ht\strutbox}{\includegraphics[height=\ht\strutbox]{figures/flags/fast_svg-tex}}%
  \endgroup} {\begingroup\normalfont
  \raisebox{-.1\ht\strutbox}{\includegraphics[height=\ht\strutbox]{figures/flags/guard_svg-tex}}%
  \endgroup}
        {\begingroup\normalfont
  \raisebox{-.1\ht\strutbox}{\includegraphics[height=\ht\strutbox]{figures/flags/hsdir_svg-tex}}%
  \endgroup} {\begingroup\normalfont
  \raisebox{-.1\ht\strutbox}{\includegraphics[height=\ht\strutbox]{figures/flags/running_svg-tex}}%
  \endgroup} {\begingroup\normalfont
  \raisebox{-.1\ht\strutbox}{\includegraphics[height=\ht\strutbox]{figures/flags/stable_svg-tex}}%
  \endgroup}
        {\begingroup\normalfont
  \raisebox{-.1\ht\strutbox}{\includegraphics[height=\ht\strutbox]{figures/flags/v2dir_svg-tex}}%
  \endgroup} {\begingroup\normalfont
  \raisebox{-.1\ht\strutbox}{\includegraphics[height=\ht\strutbox]{figures/flags/valid_svg-tex}}%
  \endgroup} \\

        32 & \textcolor{blue}{\textbf{presaultboubd8}} & \num{3093893} & RO (200 km) & \as[link=true]{47890} & 41.73~MiB/s & 2021-12-08 & 77 days &
        {\begingroup\normalfont
  \raisebox{-.1\ht\strutbox}{\includegraphics[height=\ht\strutbox]{figures/flags/exit_svg-tex}}%
  \endgroup} {\begingroup\normalfont
  \raisebox{-.1\ht\strutbox}{\includegraphics[height=\ht\strutbox]{figures/flags/fast_svg-tex}}%
  \endgroup} {\begingroup\normalfont
  \raisebox{-.1\ht\strutbox}{\includegraphics[height=\ht\strutbox]{figures/flags/guard_svg-tex}}%
  \endgroup}
        {\begingroup\normalfont
  \raisebox{-.1\ht\strutbox}{\includegraphics[height=\ht\strutbox]{figures/flags/hsdir_svg-tex}}%
  \endgroup} {\begingroup\normalfont
  \raisebox{-.1\ht\strutbox}{\includegraphics[height=\ht\strutbox]{figures/flags/running_svg-tex}}%
  \endgroup} {\begingroup\normalfont
  \raisebox{-.1\ht\strutbox}{\includegraphics[height=\ht\strutbox]{figures/flags/stable_svg-tex}}%
  \endgroup}
        {\begingroup\normalfont
  \raisebox{-.1\ht\strutbox}{\includegraphics[height=\ht\strutbox]{figures/flags/v2dir_svg-tex}}%
  \endgroup} {\begingroup\normalfont
  \raisebox{-.1\ht\strutbox}{\includegraphics[height=\ht\strutbox]{figures/flags/valid_svg-tex}}%
  \endgroup} \\

        67 & \textcolor{blue}{\textbf{smortRley}} & \num{2394694} & Uusimaa, FI (200 km) & \as[link=true]{24940} & 56.90~MiB/s & 2022-07-06 & 8 days &
        {\begingroup\normalfont
  \raisebox{-.1\ht\strutbox}{\includegraphics[height=\ht\strutbox]{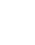}}%
  \endgroup} {\begingroup\normalfont
  \raisebox{-.1\ht\strutbox}{\includegraphics[height=\ht\strutbox]{figures/flags/fast_svg-tex}}%
  \endgroup} {\begingroup\normalfont
  \raisebox{-.1\ht\strutbox}{\includegraphics[height=\ht\strutbox]{figures/flags/guard_svg-tex}}%
  \endgroup}
        {\begingroup\normalfont
  \raisebox{-.1\ht\strutbox}{\includegraphics[height=\ht\strutbox]{figures/flags/no-hsdir_svg-tex}}%
  \endgroup} {\begingroup\normalfont
  \raisebox{-.1\ht\strutbox}{\includegraphics[height=\ht\strutbox]{figures/flags/running_svg-tex}}%
  \endgroup} {\begingroup\normalfont
  \raisebox{-.1\ht\strutbox}{\includegraphics[height=\ht\strutbox]{figures/flags/stable_svg-tex}}%
  \endgroup}
        {\begingroup\normalfont
  \raisebox{-.1\ht\strutbox}{\includegraphics[height=\ht\strutbox]{figures/flags/v2dir_svg-tex}}%
  \endgroup} {\begingroup\normalfont
  \raisebox{-.1\ht\strutbox}{\includegraphics[height=\ht\strutbox]{figures/flags/valid_svg-tex}}%
  \endgroup} \\

        177 & \textcolor{blue}{\textbf{Password}} & \num{1756601} & Bucharest, RO (20 km) & \as[link=true]{200651} & 26.45~MiB/s & 2021-12-06 & 306 days &
        {\begingroup\normalfont
  \raisebox{-.1\ht\strutbox}{\includegraphics[height=\ht\strutbox]{figures/flags/exit_svg-tex}}%
  \endgroup} {\begingroup\normalfont
  \raisebox{-.1\ht\strutbox}{\includegraphics[height=\ht\strutbox]{figures/flags/fast_svg-tex}}%
  \endgroup} {\begingroup\normalfont
  \raisebox{-.1\ht\strutbox}{\includegraphics[height=\ht\strutbox]{figures/flags/guard_svg-tex}}%
  \endgroup}
        {\begingroup\normalfont
  \raisebox{-.1\ht\strutbox}{\includegraphics[height=\ht\strutbox]{figures/flags/hsdir_svg-tex}}%
  \endgroup} {\begingroup\normalfont
  \raisebox{-.1\ht\strutbox}{\includegraphics[height=\ht\strutbox]{figures/flags/running_svg-tex}}%
  \endgroup} {\begingroup\normalfont
  \raisebox{-.1\ht\strutbox}{\includegraphics[height=\ht\strutbox]{figures/flags/stable_svg-tex}}%
  \endgroup}
        {\begingroup\normalfont
  \raisebox{-.1\ht\strutbox}{\includegraphics[height=\ht\strutbox]{figures/flags/v2dir_svg-tex}}%
  \endgroup} {\begingroup\normalfont
  \raisebox{-.1\ht\strutbox}{\includegraphics[height=\ht\strutbox]{figures/flags/valid_svg-tex}}%
  \endgroup} \\

        267 & \textcolor{blue}{\textbf{z0rb4l2}} & \num{1436816} & DE (200 km) & \as[link=true]{24940} & 45.57~MiB/s & 2017-04-22 & 105 days &
        {\begingroup\normalfont
  \raisebox{-.1\ht\strutbox}{\includegraphics[height=\ht\strutbox]{figures/flags/no-exit_svg-tex}}%
  \endgroup} {\begingroup\normalfont
  \raisebox{-.1\ht\strutbox}{\includegraphics[height=\ht\strutbox]{figures/flags/fast_svg-tex}}%
  \endgroup} {\begingroup\normalfont
  \raisebox{-.1\ht\strutbox}{\includegraphics[height=\ht\strutbox]{figures/flags/guard_svg-tex}}%
  \endgroup}
        {\begingroup\normalfont
  \raisebox{-.1\ht\strutbox}{\includegraphics[height=\ht\strutbox]{figures/flags/no-hsdir_svg-tex}}%
  \endgroup} {\begingroup\normalfont
  \raisebox{-.1\ht\strutbox}{\includegraphics[height=\ht\strutbox]{figures/flags/running_svg-tex}}%
  \endgroup} {\begingroup\normalfont
  \raisebox{-.1\ht\strutbox}{\includegraphics[height=\ht\strutbox]{figures/flags/stable_svg-tex}}%
  \endgroup}
        {\begingroup\normalfont
  \raisebox{-.1\ht\strutbox}{\includegraphics[height=\ht\strutbox]{figures/flags/v2dir_svg-tex}}%
  \endgroup} {\begingroup\normalfont
  \raisebox{-.1\ht\strutbox}{\includegraphics[height=\ht\strutbox]{figures/flags/valid_svg-tex}}%
  \endgroup} \\

        276 & \textcolor{blue}{\textbf{WWW}} & \num{1419711} & FR (500 km) & \as[link=true]{16276} & 46.06~MiB/s & 2021-08-24 & 10 days &
        {\begingroup\normalfont
  \raisebox{-.1\ht\strutbox}{\includegraphics[height=\ht\strutbox]{figures/flags/no-exit_svg-tex}}%
  \endgroup} {\begingroup\normalfont
  \raisebox{-.1\ht\strutbox}{\includegraphics[height=\ht\strutbox]{figures/flags/fast_svg-tex}}%
  \endgroup} {\begingroup\normalfont
  \raisebox{-.1\ht\strutbox}{\includegraphics[height=\ht\strutbox]{figures/flags/guard_svg-tex}}%
  \endgroup}
        {\begingroup\normalfont
  \raisebox{-.1\ht\strutbox}{\includegraphics[height=\ht\strutbox]{figures/flags/hsdir_svg-tex}}%
  \endgroup} {\begingroup\normalfont
  \raisebox{-.1\ht\strutbox}{\includegraphics[height=\ht\strutbox]{figures/flags/running_svg-tex}}%
  \endgroup} {\begingroup\normalfont
  \raisebox{-.1\ht\strutbox}{\includegraphics[height=\ht\strutbox]{figures/flags/stable_svg-tex}}%
  \endgroup}
        {\begingroup\normalfont
  \raisebox{-.1\ht\strutbox}{\includegraphics[height=\ht\strutbox]{figures/flags/v2dir_svg-tex}}%
  \endgroup} {\begingroup\normalfont
  \raisebox{-.1\ht\strutbox}{\includegraphics[height=\ht\strutbox]{figures/flags/valid_svg-tex}}%
  \endgroup} \\

        444 & \textcolor{blue}{\textbf{lightblue}} & \num{1128927} & DE (200 km) & \as[link=true]{42730} & 34.70~MiB/s & 2020-08-15 & 36 days &
        {\begingroup\normalfont
  \raisebox{-.1\ht\strutbox}{\includegraphics[height=\ht\strutbox]{figures/flags/no-exit_svg-tex}}%
  \endgroup} {\begingroup\normalfont
  \raisebox{-.1\ht\strutbox}{\includegraphics[height=\ht\strutbox]{figures/flags/fast_svg-tex}}%
  \endgroup} {\begingroup\normalfont
  \raisebox{-.1\ht\strutbox}{\includegraphics[height=\ht\strutbox]{figures/flags/guard_svg-tex}}%
  \endgroup}
        {\begingroup\normalfont
  \raisebox{-.1\ht\strutbox}{\includegraphics[height=\ht\strutbox]{figures/flags/no-hsdir_svg-tex}}%
  \endgroup} {\begingroup\normalfont
  \raisebox{-.1\ht\strutbox}{\includegraphics[height=\ht\strutbox]{figures/flags/running_svg-tex}}%
  \endgroup} {\begingroup\normalfont
  \raisebox{-.1\ht\strutbox}{\includegraphics[height=\ht\strutbox]{figures/flags/stable_svg-tex}}%
  \endgroup}
        {\begingroup\normalfont
  \raisebox{-.1\ht\strutbox}{\includegraphics[height=\ht\strutbox]{figures/flags/v2dir_svg-tex}}%
  \endgroup} {\begingroup\normalfont
  \raisebox{-.1\ht\strutbox}{\includegraphics[height=\ht\strutbox]{figures/flags/valid_svg-tex}}%
  \endgroup} \\

        571 & \textcolor{blue}{\textbf{CanisFamiliaris}}~({\begingroup\normalfont
  \raisebox{-.1\ht\strutbox}{\includegraphics[height=\ht\strutbox]{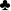}}%
  \endgroup}) & \num{974982} & US (1000 km) & \as[link=true]{16276} & 42.32~MiB/s & 2022-03-23 & 1 day\phantom{s} &
        {\begingroup\normalfont
  \raisebox{-.1\ht\strutbox}{\includegraphics[height=\ht\strutbox]{figures/flags/no-exit_svg-tex}}%
  \endgroup} {\begingroup\normalfont
  \raisebox{-.1\ht\strutbox}{\includegraphics[height=\ht\strutbox]{figures/flags/fast_svg-tex}}%
  \endgroup} {\begingroup\normalfont
  \raisebox{-.1\ht\strutbox}{\includegraphics[height=\ht\strutbox]{figures/flags/guard_svg-tex}}%
  \endgroup}
        {\begingroup\normalfont
  \raisebox{-.1\ht\strutbox}{\includegraphics[height=\ht\strutbox]{figures/flags/no-hsdir_svg-tex}}%
  \endgroup} {\begingroup\normalfont
  \raisebox{-.1\ht\strutbox}{\includegraphics[height=\ht\strutbox]{figures/flags/running_svg-tex}}%
  \endgroup} {\begingroup\normalfont
  \raisebox{-.1\ht\strutbox}{\includegraphics[height=\ht\strutbox]{figures/flags/stable_svg-tex}}%
  \endgroup}
        {\begingroup\normalfont
  \raisebox{-.1\ht\strutbox}{\includegraphics[height=\ht\strutbox]{figures/flags/v2dir_svg-tex}}%
  \endgroup} {\begingroup\normalfont
  \raisebox{-.1\ht\strutbox}{\includegraphics[height=\ht\strutbox]{figures/flags/valid_svg-tex}}%
  \endgroup} \\

        608 & \textcolor{blue}{\textbf{Euphoria}} & \num{940772} & Virginia, US (1000 km) & \as[link=true]{16276} & 40.02~MiB/s & 2022-03-20 & 2 days &
        {\begingroup\normalfont
  \raisebox{-.1\ht\strutbox}{\includegraphics[height=\ht\strutbox]{figures/flags/no-exit_svg-tex}}%
  \endgroup} {\begingroup\normalfont
  \raisebox{-.1\ht\strutbox}{\includegraphics[height=\ht\strutbox]{figures/flags/fast_svg-tex}}%
  \endgroup} {\begingroup\normalfont
  \raisebox{-.1\ht\strutbox}{\includegraphics[height=\ht\strutbox]{figures/flags/guard_svg-tex}}%
  \endgroup}
        {\begingroup\normalfont
  \raisebox{-.1\ht\strutbox}{\includegraphics[height=\ht\strutbox]{figures/flags/no-hsdir_svg-tex}}%
  \endgroup} {\begingroup\normalfont
  \raisebox{-.1\ht\strutbox}{\includegraphics[height=\ht\strutbox]{figures/flags/running_svg-tex}}%
  \endgroup} {\begingroup\normalfont
  \raisebox{-.1\ht\strutbox}{\includegraphics[height=\ht\strutbox]{figures/flags/stable_svg-tex}}%
  \endgroup}
        {\begingroup\normalfont
  \raisebox{-.1\ht\strutbox}{\includegraphics[height=\ht\strutbox]{figures/flags/v2dir_svg-tex}}%
  \endgroup} {\begingroup\normalfont
  \raisebox{-.1\ht\strutbox}{\includegraphics[height=\ht\strutbox]{figures/flags/valid_svg-tex}}%
  \endgroup} \\

        621 & \textcolor{blue}{\textbf{CanisLupus}}~({\begingroup\normalfont
  \raisebox{-.1\ht\strutbox}{\includegraphics[height=\ht\strutbox]{figures/clubsuit_svg-tex}}%
  \endgroup}) & \num{923667} & US (1000 km) & \as[link=true]{16276} & 40.90~MiB/s & 2022-03-23 & 1 day\phantom{s} &
        {\begingroup\normalfont
  \raisebox{-.1\ht\strutbox}{\includegraphics[height=\ht\strutbox]{figures/flags/no-exit_svg-tex}}%
  \endgroup} {\begingroup\normalfont
  \raisebox{-.1\ht\strutbox}{\includegraphics[height=\ht\strutbox]{figures/flags/fast_svg-tex}}%
  \endgroup} {\begingroup\normalfont
  \raisebox{-.1\ht\strutbox}{\includegraphics[height=\ht\strutbox]{figures/flags/guard_svg-tex}}%
  \endgroup}
        {\begingroup\normalfont
  \raisebox{-.1\ht\strutbox}{\includegraphics[height=\ht\strutbox]{figures/flags/hsdir_svg-tex}}%
  \endgroup} {\begingroup\normalfont
  \raisebox{-.1\ht\strutbox}{\includegraphics[height=\ht\strutbox]{figures/flags/running_svg-tex}}%
  \endgroup} {\begingroup\normalfont
  \raisebox{-.1\ht\strutbox}{\includegraphics[height=\ht\strutbox]{figures/flags/stable_svg-tex}}%
  \endgroup}
        {\begingroup\normalfont
  \raisebox{-.1\ht\strutbox}{\includegraphics[height=\ht\strutbox]{figures/flags/v2dir_svg-tex}}%
  \endgroup} {\begingroup\normalfont
  \raisebox{-.1\ht\strutbox}{\includegraphics[height=\ht\strutbox]{figures/flags/valid_svg-tex}}%
  \endgroup} \\

        627 & \textcolor{blue}{\textbf{adrian}} & \num{923667} & Virginia, US (1000 km) & \as[link=true]{16276} & 44.83~MiB/s & 2022-01-26 & 48 days &
        {\begingroup\normalfont
  \raisebox{-.1\ht\strutbox}{\includegraphics[height=\ht\strutbox]{figures/flags/no-exit_svg-tex}}%
  \endgroup} {\begingroup\normalfont
  \raisebox{-.1\ht\strutbox}{\includegraphics[height=\ht\strutbox]{figures/flags/fast_svg-tex}}%
  \endgroup} {\begingroup\normalfont
  \raisebox{-.1\ht\strutbox}{\includegraphics[height=\ht\strutbox]{figures/flags/guard_svg-tex}}%
  \endgroup}
        {\begingroup\normalfont
  \raisebox{-.1\ht\strutbox}{\includegraphics[height=\ht\strutbox]{figures/flags/hsdir_svg-tex}}%
  \endgroup} {\begingroup\normalfont
  \raisebox{-.1\ht\strutbox}{\includegraphics[height=\ht\strutbox]{figures/flags/running_svg-tex}}%
  \endgroup} {\begingroup\normalfont
  \raisebox{-.1\ht\strutbox}{\includegraphics[height=\ht\strutbox]{figures/flags/stable_svg-tex}}%
  \endgroup}
        {\begingroup\normalfont
  \raisebox{-.1\ht\strutbox}{\includegraphics[height=\ht\strutbox]{figures/flags/v2dir_svg-tex}}%
  \endgroup} {\begingroup\normalfont
  \raisebox{-.1\ht\strutbox}{\includegraphics[height=\ht\strutbox]{figures/flags/valid_svg-tex}}%
  \endgroup} \\

        669 & \textcolor{blue}{\textbf{Bananenbakker}} & \num{902919} & NL (100 km) & \as[link=true]{1102} & 37.86~MiB/s & 2018-05-04 & 5 days &
        {\begingroup\normalfont
  \raisebox{-.1\ht\strutbox}{\includegraphics[height=\ht\strutbox]{figures/flags/no-exit_svg-tex}}%
  \endgroup} {\begingroup\normalfont
  \raisebox{-.1\ht\strutbox}{\includegraphics[height=\ht\strutbox]{figures/flags/fast_svg-tex}}%
  \endgroup} {\begingroup\normalfont
  \raisebox{-.1\ht\strutbox}{\includegraphics[height=\ht\strutbox]{figures/flags/guard_svg-tex}}%
  \endgroup}
        {\begingroup\normalfont
  \raisebox{-.1\ht\strutbox}{\includegraphics[height=\ht\strutbox]{figures/flags/no-hsdir_svg-tex}}%
  \endgroup} {\begingroup\normalfont
  \raisebox{-.1\ht\strutbox}{\includegraphics[height=\ht\strutbox]{figures/flags/running_svg-tex}}%
  \endgroup} {\begingroup\normalfont
  \raisebox{-.1\ht\strutbox}{\includegraphics[height=\ht\strutbox]{figures/flags/stable_svg-tex}}%
  \endgroup}
        {\begingroup\normalfont
  \raisebox{-.1\ht\strutbox}{\includegraphics[height=\ht\strutbox]{figures/flags/v2dir_svg-tex}}%
  \endgroup} {\begingroup\normalfont
  \raisebox{-.1\ht\strutbox}{\includegraphics[height=\ht\strutbox]{figures/flags/valid_svg-tex}}%
  \endgroup} \\

        711 & torx1steack & \num{855248} & Lyon, FR (20 km) & \as[link=true]{207992} & 40.90~MiB/s & 2021-12-04 & 2 days &
        {\begingroup\normalfont
  \raisebox{-.1\ht\strutbox}{\includegraphics[height=\ht\strutbox]{figures/flags/no-exit_svg-tex}}%
  \endgroup} {\begingroup\normalfont
  \raisebox{-.1\ht\strutbox}{\includegraphics[height=\ht\strutbox]{figures/flags/fast_svg-tex}}%
  \endgroup} {\begingroup\normalfont
  \raisebox{-.1\ht\strutbox}{\includegraphics[height=\ht\strutbox]{figures/flags/guard_svg-tex}}%
  \endgroup}
        {\begingroup\normalfont
  \raisebox{-.1\ht\strutbox}{\includegraphics[height=\ht\strutbox]{figures/flags/no-hsdir_svg-tex}}%
  \endgroup} {\begingroup\normalfont
  \raisebox{-.1\ht\strutbox}{\includegraphics[height=\ht\strutbox]{figures/flags/running_svg-tex}}%
  \endgroup} {\begingroup\normalfont
  \raisebox{-.1\ht\strutbox}{\includegraphics[height=\ht\strutbox]{figures/flags/stable_svg-tex}}%
  \endgroup}
        {\begingroup\normalfont
  \raisebox{-.1\ht\strutbox}{\includegraphics[height=\ht\strutbox]{figures/flags/v2dir_svg-tex}}%
  \endgroup} {\begingroup\normalfont
  \raisebox{-.1\ht\strutbox}{\includegraphics[height=\ht\strutbox]{figures/flags/valid_svg-tex}}%
  \endgroup} \\

        \bottomrule
    \end{tabular}
    \begin{tablenotes}[flushleft]
        \item[a] To improve readability of the metric value, we multiply it by a factor of $10^9$.
        \label{note:table-1}
    \end{tablenotes}
    \end{threeparttable}
    \caption{Top-15 relays ranked by relay utility metric ($RM$)}
    \label{tab:relay_top15}
\end{subtable}

    \quad  

    \begin{subtable}{\textwidth}
        \centering
    \begin{tabular}{rlrllS[table-format=-1.2,table-number-alignment = left]lrl}
        \toprule
        \textbf{\#} & \textbf{Relay (family)} & \textbf{$NM$} & \textbf{Location (accuracy)} & \textbf{\ac{as}} &
        \textbf{Bandwidth} & \textbf{First seen} & \textbf{Uptime} & \textbf{Flags}
        \\
        \midrule
        2 & \textcolor{blue}{\textbf{slauthtertodeport1}} & \num{1693736} & LT (200 km) & \as[link=true]{209588} & 75.31~MiB/s & 2021-12-08 & 93 days &
        {\begingroup\normalfont
  \raisebox{-.1\ht\strutbox}{\includegraphics[height=\ht\strutbox]{figures/flags/exit_svg-tex}}%
  \endgroup} {\begingroup\normalfont
  \raisebox{-.1\ht\strutbox}{\includegraphics[height=\ht\strutbox]{figures/flags/fast_svg-tex}}%
  \endgroup} {\begingroup\normalfont
  \raisebox{-.1\ht\strutbox}{\includegraphics[height=\ht\strutbox]{figures/flags/guard_svg-tex}}%
  \endgroup}
        {\begingroup\normalfont
  \raisebox{-.1\ht\strutbox}{\includegraphics[height=\ht\strutbox]{figures/flags/hsdir_svg-tex}}%
  \endgroup} {\begingroup\normalfont
  \raisebox{-.1\ht\strutbox}{\includegraphics[height=\ht\strutbox]{figures/flags/running_svg-tex}}%
  \endgroup} {\begingroup\normalfont
  \raisebox{-.1\ht\strutbox}{\includegraphics[height=\ht\strutbox]{figures/flags/stable_svg-tex}}%
  \endgroup}
        {\begingroup\normalfont
  \raisebox{-.1\ht\strutbox}{\includegraphics[height=\ht\strutbox]{figures/flags/v2dir_svg-tex}}%
  \endgroup} {\begingroup\normalfont
  \raisebox{-.1\ht\strutbox}{\includegraphics[height=\ht\strutbox]{figures/flags/valid_svg-tex}}%
  \endgroup} \\

        5 & \textcolor{blue}{\textbf{NLfreedom1}} & \num{1184356} & Dronten, NL (20 km) & \as[link=true]{60404} & 68.20~MiB/s & 2022-02-27 & 222 days &
        {\begingroup\normalfont
  \raisebox{-.1\ht\strutbox}{\includegraphics[height=\ht\strutbox]{figures/flags/exit_svg-tex}}%
  \endgroup} {\begingroup\normalfont
  \raisebox{-.1\ht\strutbox}{\includegraphics[height=\ht\strutbox]{figures/flags/fast_svg-tex}}%
  \endgroup} {\begingroup\normalfont
  \raisebox{-.1\ht\strutbox}{\includegraphics[height=\ht\strutbox]{figures/flags/guard_svg-tex}}%
  \endgroup}
        {\begingroup\normalfont
  \raisebox{-.1\ht\strutbox}{\includegraphics[height=\ht\strutbox]{figures/flags/hsdir_svg-tex}}%
  \endgroup} {\begingroup\normalfont
  \raisebox{-.1\ht\strutbox}{\includegraphics[height=\ht\strutbox]{figures/flags/running_svg-tex}}%
  \endgroup} {\begingroup\normalfont
  \raisebox{-.1\ht\strutbox}{\includegraphics[height=\ht\strutbox]{figures/flags/stable_svg-tex}}%
  \endgroup}
        {\begingroup\normalfont
  \raisebox{-.1\ht\strutbox}{\includegraphics[height=\ht\strutbox]{figures/flags/v2dir_svg-tex}}%
  \endgroup} {\begingroup\normalfont
  \raisebox{-.1\ht\strutbox}{\includegraphics[height=\ht\strutbox]{figures/flags/valid_svg-tex}}%
  \endgroup} \\

        9 & \textcolor{blue}{\textbf{xeracition}} & \num{1118268} & DE (200 km) & \as[link=true]{51862} & 47.11~MiB/s & 2022-09-08 & 2 days &
        {\begingroup\normalfont
  \raisebox{-.1\ht\strutbox}{\includegraphics[height=\ht\strutbox]{figures/flags/exit_svg-tex}}%
  \endgroup} {\begingroup\normalfont
  \raisebox{-.1\ht\strutbox}{\includegraphics[height=\ht\strutbox]{figures/flags/fast_svg-tex}}%
  \endgroup} {\begingroup\normalfont
  \raisebox{-.1\ht\strutbox}{\includegraphics[height=\ht\strutbox]{figures/flags/guard_svg-tex}}%
  \endgroup}
        {\begingroup\normalfont
  \raisebox{-.1\ht\strutbox}{\includegraphics[height=\ht\strutbox]{figures/flags/no-hsdir_svg-tex}}%
  \endgroup} {\begingroup\normalfont
  \raisebox{-.1\ht\strutbox}{\includegraphics[height=\ht\strutbox]{figures/flags/running_svg-tex}}%
  \endgroup} {\begingroup\normalfont
  \raisebox{-.1\ht\strutbox}{\includegraphics[height=\ht\strutbox]{figures/flags/stable_svg-tex}}%
  \endgroup}
        {\begingroup\normalfont
  \raisebox{-.1\ht\strutbox}{\includegraphics[height=\ht\strutbox]{figures/flags/v2dir_svg-tex}}%
  \endgroup} {\begingroup\normalfont
  \raisebox{-.1\ht\strutbox}{\includegraphics[height=\ht\strutbox]{figures/flags/valid_svg-tex}}%
  \endgroup} \\

        31 & \textcolor{blue}{\textbf{smortRley}} & \num{869508} & Uusimaa, FI (200 km) & \as[link=true]{24940} & 56.90~MiB/s & 2022-07-06 & 8 days &
        {\begingroup\normalfont
  \raisebox{-.1\ht\strutbox}{\includegraphics[height=\ht\strutbox]{figures/flags/no-exit_svg-tex}}%
  \endgroup} {\begingroup\normalfont
  \raisebox{-.1\ht\strutbox}{\includegraphics[height=\ht\strutbox]{figures/flags/fast_svg-tex}}%
  \endgroup} {\begingroup\normalfont
  \raisebox{-.1\ht\strutbox}{\includegraphics[height=\ht\strutbox]{figures/flags/guard_svg-tex}}%
  \endgroup}
        {\begingroup\normalfont
  \raisebox{-.1\ht\strutbox}{\includegraphics[height=\ht\strutbox]{figures/flags/no-hsdir_svg-tex}}%
  \endgroup} {\begingroup\normalfont
  \raisebox{-.1\ht\strutbox}{\includegraphics[height=\ht\strutbox]{figures/flags/running_svg-tex}}%
  \endgroup} {\begingroup\normalfont
  \raisebox{-.1\ht\strutbox}{\includegraphics[height=\ht\strutbox]{figures/flags/stable_svg-tex}}%
  \endgroup}
        {\begingroup\normalfont
  \raisebox{-.1\ht\strutbox}{\includegraphics[height=\ht\strutbox]{figures/flags/v2dir_svg-tex}}%
  \endgroup} {\begingroup\normalfont
  \raisebox{-.1\ht\strutbox}{\includegraphics[height=\ht\strutbox]{figures/flags/valid_svg-tex}}%
  \endgroup} \\

        62 & \textcolor{blue}{\textbf{presaultboubd8}} & \num{714523} & RO (200 km) & \as[link=true]{47890} & 41.73~MiB/s & 2021-12-08 & 77 days &
        {\begingroup\normalfont
  \raisebox{-.1\ht\strutbox}{\includegraphics[height=\ht\strutbox]{figures/flags/exit_svg-tex}}%
  \endgroup} {\begingroup\normalfont
  \raisebox{-.1\ht\strutbox}{\includegraphics[height=\ht\strutbox]{figures/flags/fast_svg-tex}}%
  \endgroup} {\begingroup\normalfont
  \raisebox{-.1\ht\strutbox}{\includegraphics[height=\ht\strutbox]{figures/flags/guard_svg-tex}}%
  \endgroup}
        {\begingroup\normalfont
  \raisebox{-.1\ht\strutbox}{\includegraphics[height=\ht\strutbox]{figures/flags/hsdir_svg-tex}}%
  \endgroup} {\begingroup\normalfont
  \raisebox{-.1\ht\strutbox}{\includegraphics[height=\ht\strutbox]{figures/flags/running_svg-tex}}%
  \endgroup} {\begingroup\normalfont
  \raisebox{-.1\ht\strutbox}{\includegraphics[height=\ht\strutbox]{figures/flags/stable_svg-tex}}%
  \endgroup}
        {\begingroup\normalfont
  \raisebox{-.1\ht\strutbox}{\includegraphics[height=\ht\strutbox]{figures/flags/v2dir_svg-tex}}%
  \endgroup} {\begingroup\normalfont
  \raisebox{-.1\ht\strutbox}{\includegraphics[height=\ht\strutbox]{figures/flags/valid_svg-tex}}%
  \endgroup} \\

        162 & \textcolor{blue}{\textbf{z0rb4l2}} & \num{521705} & DE (200 km) & \as[link=true]{24940} & 45.57~MiB/s & 2017-04-22 & 105 days &
        {\begingroup\normalfont
  \raisebox{-.1\ht\strutbox}{\includegraphics[height=\ht\strutbox]{figures/flags/no-exit_svg-tex}}%
  \endgroup} {\begingroup\normalfont
  \raisebox{-.1\ht\strutbox}{\includegraphics[height=\ht\strutbox]{figures/flags/fast_svg-tex}}%
  \endgroup} {\begingroup\normalfont
  \raisebox{-.1\ht\strutbox}{\includegraphics[height=\ht\strutbox]{figures/flags/guard_svg-tex}}%
  \endgroup}
        {\begingroup\normalfont
  \raisebox{-.1\ht\strutbox}{\includegraphics[height=\ht\strutbox]{figures/flags/no-hsdir_svg-tex}}%
  \endgroup} {\begingroup\normalfont
  \raisebox{-.1\ht\strutbox}{\includegraphics[height=\ht\strutbox]{figures/flags/running_svg-tex}}%
  \endgroup} {\begingroup\normalfont
  \raisebox{-.1\ht\strutbox}{\includegraphics[height=\ht\strutbox]{figures/flags/stable_svg-tex}}%
  \endgroup}
        {\begingroup\normalfont
  \raisebox{-.1\ht\strutbox}{\includegraphics[height=\ht\strutbox]{figures/flags/v2dir_svg-tex}}%
  \endgroup} {\begingroup\normalfont
  \raisebox{-.1\ht\strutbox}{\includegraphics[height=\ht\strutbox]{figures/flags/valid_svg-tex}}%
  \endgroup} \\

        178 & \textcolor{blue}{\textbf{WWW}} & \num{506756} & FR (500 km) & \as[link=true]{16276} & 46.06~MiB/s & 2021-08-24 & 10 days &
        {\begingroup\normalfont
  \raisebox{-.1\ht\strutbox}{\includegraphics[height=\ht\strutbox]{figures/flags/no-exit_svg-tex}}%
  \endgroup} {\begingroup\normalfont
  \raisebox{-.1\ht\strutbox}{\includegraphics[height=\ht\strutbox]{figures/flags/fast_svg-tex}}%
  \endgroup} {\begingroup\normalfont
  \raisebox{-.1\ht\strutbox}{\includegraphics[height=\ht\strutbox]{figures/flags/guard_svg-tex}}%
  \endgroup}
        {\begingroup\normalfont
  \raisebox{-.1\ht\strutbox}{\includegraphics[height=\ht\strutbox]{figures/flags/hsdir_svg-tex}}%
  \endgroup} {\begingroup\normalfont
  \raisebox{-.1\ht\strutbox}{\includegraphics[height=\ht\strutbox]{figures/flags/running_svg-tex}}%
  \endgroup} {\begingroup\normalfont
  \raisebox{-.1\ht\strutbox}{\includegraphics[height=\ht\strutbox]{figures/flags/stable_svg-tex}}%
  \endgroup}
        {\begingroup\normalfont
  \raisebox{-.1\ht\strutbox}{\includegraphics[height=\ht\strutbox]{figures/flags/v2dir_svg-tex}}%
  \endgroup} {\begingroup\normalfont
  \raisebox{-.1\ht\strutbox}{\includegraphics[height=\ht\strutbox]{figures/flags/valid_svg-tex}}%
  \endgroup} \\

        338 & \textcolor{blue}{\textbf{lightblue}} & \num{384364} & DE (200 km) & \as[link=true]{42730} & 34.70~MiB/s & 2020-08-15 & 36 days &
        {\begingroup\normalfont
  \raisebox{-.1\ht\strutbox}{\includegraphics[height=\ht\strutbox]{figures/flags/no-exit_svg-tex}}%
  \endgroup} {\begingroup\normalfont
  \raisebox{-.1\ht\strutbox}{\includegraphics[height=\ht\strutbox]{figures/flags/fast_svg-tex}}%
  \endgroup} {\begingroup\normalfont
  \raisebox{-.1\ht\strutbox}{\includegraphics[height=\ht\strutbox]{figures/flags/guard_svg-tex}}%
  \endgroup}
        {\begingroup\normalfont
  \raisebox{-.1\ht\strutbox}{\includegraphics[height=\ht\strutbox]{figures/flags/no-hsdir_svg-tex}}%
  \endgroup} {\begingroup\normalfont
  \raisebox{-.1\ht\strutbox}{\includegraphics[height=\ht\strutbox]{figures/flags/running_svg-tex}}%
  \endgroup} {\begingroup\normalfont
  \raisebox{-.1\ht\strutbox}{\includegraphics[height=\ht\strutbox]{figures/flags/stable_svg-tex}}%
  \endgroup}
        {\begingroup\normalfont
  \raisebox{-.1\ht\strutbox}{\includegraphics[height=\ht\strutbox]{figures/flags/v2dir_svg-tex}}%
  \endgroup} {\begingroup\normalfont
  \raisebox{-.1\ht\strutbox}{\includegraphics[height=\ht\strutbox]{figures/flags/valid_svg-tex}}%
  \endgroup} \\

        365 & \textcolor{blue}{\textbf{Password}} & \num{366237} & Bucharest, RO (20 km) & \as[link=true]{200651} & 26.45~MiB/s & 2021-12-06 & 306 days &
        {\begingroup\normalfont
  \raisebox{-.1\ht\strutbox}{\includegraphics[height=\ht\strutbox]{figures/flags/exit_svg-tex}}%
  \endgroup} {\begingroup\normalfont
  \raisebox{-.1\ht\strutbox}{\includegraphics[height=\ht\strutbox]{figures/flags/fast_svg-tex}}%
  \endgroup} {\begingroup\normalfont
  \raisebox{-.1\ht\strutbox}{\includegraphics[height=\ht\strutbox]{figures/flags/guard_svg-tex}}%
  \endgroup}
        {\begingroup\normalfont
  \raisebox{-.1\ht\strutbox}{\includegraphics[height=\ht\strutbox]{figures/flags/hsdir_svg-tex}}%
  \endgroup} {\begingroup\normalfont
  \raisebox{-.1\ht\strutbox}{\includegraphics[height=\ht\strutbox]{figures/flags/running_svg-tex}}%
  \endgroup} {\begingroup\normalfont
  \raisebox{-.1\ht\strutbox}{\includegraphics[height=\ht\strutbox]{figures/flags/stable_svg-tex}}%
  \endgroup}
        {\begingroup\normalfont
  \raisebox{-.1\ht\strutbox}{\includegraphics[height=\ht\strutbox]{figures/flags/v2dir_svg-tex}}%
  \endgroup} {\begingroup\normalfont
  \raisebox{-.1\ht\strutbox}{\includegraphics[height=\ht\strutbox]{figures/flags/valid_svg-tex}}%
  \endgroup} \\

        403 & \textcolor{blue}{\textbf{CanisFamiliaris}}~({\begingroup\normalfont
  \raisebox{-.1\ht\strutbox}{\includegraphics[height=\ht\strutbox]{figures/clubsuit_svg-tex}}%
  \endgroup}) & \num{348013} & US (1000 km) & \as[link=true]{16276} & 42.32~MiB/s & 2022-03-23 & 1 day\phantom{s} &
        {\begingroup\normalfont
  \raisebox{-.1\ht\strutbox}{\includegraphics[height=\ht\strutbox]{figures/flags/no-exit_svg-tex}}%
  \endgroup} {\begingroup\normalfont
  \raisebox{-.1\ht\strutbox}{\includegraphics[height=\ht\strutbox]{figures/flags/fast_svg-tex}}%
  \endgroup} {\begingroup\normalfont
  \raisebox{-.1\ht\strutbox}{\includegraphics[height=\ht\strutbox]{figures/flags/guard_svg-tex}}%
  \endgroup}
        {\begingroup\normalfont
  \raisebox{-.1\ht\strutbox}{\includegraphics[height=\ht\strutbox]{figures/flags/no-hsdir_svg-tex}}%
  \endgroup} {\begingroup\normalfont
  \raisebox{-.1\ht\strutbox}{\includegraphics[height=\ht\strutbox]{figures/flags/running_svg-tex}}%
  \endgroup} {\begingroup\normalfont
  \raisebox{-.1\ht\strutbox}{\includegraphics[height=\ht\strutbox]{figures/flags/stable_svg-tex}}%
  \endgroup}
        {\begingroup\normalfont
  \raisebox{-.1\ht\strutbox}{\includegraphics[height=\ht\strutbox]{figures/flags/v2dir_svg-tex}}%
  \endgroup} {\begingroup\normalfont
  \raisebox{-.1\ht\strutbox}{\includegraphics[height=\ht\strutbox]{figures/flags/valid_svg-tex}}%
  \endgroup} \\

        431 & \textcolor{blue}{\textbf{Euphoria}} & \num{335802} & Virginia, US (1000 km) & \as[link=true]{16276} & 40.02~MiB/s & 2022-03-20 & 2 days &
        {\begingroup\normalfont
  \raisebox{-.1\ht\strutbox}{\includegraphics[height=\ht\strutbox]{figures/flags/no-exit_svg-tex}}%
  \endgroup} {\begingroup\normalfont
  \raisebox{-.1\ht\strutbox}{\includegraphics[height=\ht\strutbox]{figures/flags/fast_svg-tex}}%
  \endgroup} {\begingroup\normalfont
  \raisebox{-.1\ht\strutbox}{\includegraphics[height=\ht\strutbox]{figures/flags/guard_svg-tex}}%
  \endgroup}
        {\begingroup\normalfont
  \raisebox{-.1\ht\strutbox}{\includegraphics[height=\ht\strutbox]{figures/flags/no-hsdir_svg-tex}}%
  \endgroup} {\begingroup\normalfont
  \raisebox{-.1\ht\strutbox}{\includegraphics[height=\ht\strutbox]{figures/flags/running_svg-tex}}%
  \endgroup} {\begingroup\normalfont
  \raisebox{-.1\ht\strutbox}{\includegraphics[height=\ht\strutbox]{figures/flags/stable_svg-tex}}%
  \endgroup}
        {\begingroup\normalfont
  \raisebox{-.1\ht\strutbox}{\includegraphics[height=\ht\strutbox]{figures/flags/v2dir_svg-tex}}%
  \endgroup} {\begingroup\normalfont
  \raisebox{-.1\ht\strutbox}{\includegraphics[height=\ht\strutbox]{figures/flags/valid_svg-tex}}%
  \endgroup} \\

        443 & \textcolor{blue}{\textbf{CanisLupus}}~({\begingroup\normalfont
  \raisebox{-.1\ht\strutbox}{\includegraphics[height=\ht\strutbox]{figures/clubsuit_svg-tex}}%
  \endgroup}) & \num{329696} & US (1000 km) & \as[link=true]{16276} & 40.90~MiB/s & 2022-03-23 & 1 day\phantom{s} &
        {\begingroup\normalfont
  \raisebox{-.1\ht\strutbox}{\includegraphics[height=\ht\strutbox]{figures/flags/no-exit_svg-tex}}%
  \endgroup} {\begingroup\normalfont
  \raisebox{-.1\ht\strutbox}{\includegraphics[height=\ht\strutbox]{figures/flags/fast_svg-tex}}%
  \endgroup} {\begingroup\normalfont
  \raisebox{-.1\ht\strutbox}{\includegraphics[height=\ht\strutbox]{figures/flags/guard_svg-tex}}%
  \endgroup}
        {\begingroup\normalfont
  \raisebox{-.1\ht\strutbox}{\includegraphics[height=\ht\strutbox]{figures/flags/hsdir_svg-tex}}%
  \endgroup} {\begingroup\normalfont
  \raisebox{-.1\ht\strutbox}{\includegraphics[height=\ht\strutbox]{figures/flags/running_svg-tex}}%
  \endgroup} {\begingroup\normalfont
  \raisebox{-.1\ht\strutbox}{\includegraphics[height=\ht\strutbox]{figures/flags/stable_svg-tex}}%
  \endgroup}
        {\begingroup\normalfont
  \raisebox{-.1\ht\strutbox}{\includegraphics[height=\ht\strutbox]{figures/flags/v2dir_svg-tex}}%
  \endgroup} {\begingroup\normalfont
  \raisebox{-.1\ht\strutbox}{\includegraphics[height=\ht\strutbox]{figures/flags/valid_svg-tex}}%
  \endgroup} \\

        444 & \textcolor{blue}{\textbf{adrian}} & \num{329696} & Virginia, US (1000 km) & \as[link=true]{16276} & 44.83~MiB/s & 2022-01-26 & 48 days &
        {\begingroup\normalfont
  \raisebox{-.1\ht\strutbox}{\includegraphics[height=\ht\strutbox]{figures/flags/no-exit_svg-tex}}%
  \endgroup} {\begingroup\normalfont
  \raisebox{-.1\ht\strutbox}{\includegraphics[height=\ht\strutbox]{figures/flags/fast_svg-tex}}%
  \endgroup} {\begingroup\normalfont
  \raisebox{-.1\ht\strutbox}{\includegraphics[height=\ht\strutbox]{figures/flags/guard_svg-tex}}%
  \endgroup}
        {\begingroup\normalfont
  \raisebox{-.1\ht\strutbox}{\includegraphics[height=\ht\strutbox]{figures/flags/hsdir_svg-tex}}%
  \endgroup} {\begingroup\normalfont
  \raisebox{-.1\ht\strutbox}{\includegraphics[height=\ht\strutbox]{figures/flags/running_svg-tex}}%
  \endgroup} {\begingroup\normalfont
  \raisebox{-.1\ht\strutbox}{\includegraphics[height=\ht\strutbox]{figures/flags/stable_svg-tex}}%
  \endgroup}
        {\begingroup\normalfont
  \raisebox{-.1\ht\strutbox}{\includegraphics[height=\ht\strutbox]{figures/flags/v2dir_svg-tex}}%
  \endgroup} {\begingroup\normalfont
  \raisebox{-.1\ht\strutbox}{\includegraphics[height=\ht\strutbox]{figures/flags/valid_svg-tex}}%
  \endgroup} \\

        483 & Unnamed & \num{313283} & Fremont, US (20 km) & \as[link=true]{63949} & 32.12~MiB/s & 2022-09-24 & 94 days &
        {\begingroup\normalfont
  \raisebox{-.1\ht\strutbox}{\includegraphics[height=\ht\strutbox]{figures/flags/no-exit_svg-tex}}%
  \endgroup} {\begingroup\normalfont
  \raisebox{-.1\ht\strutbox}{\includegraphics[height=\ht\strutbox]{figures/flags/fast_svg-tex}}%
  \endgroup} {\begingroup\normalfont
  \raisebox{-.1\ht\strutbox}{\includegraphics[height=\ht\strutbox]{figures/flags/guard_svg-tex}}%
  \endgroup}
        {\begingroup\normalfont
  \raisebox{-.1\ht\strutbox}{\includegraphics[height=\ht\strutbox]{figures/flags/no-hsdir_svg-tex}}%
  \endgroup} {\begingroup\normalfont
  \raisebox{-.1\ht\strutbox}{\includegraphics[height=\ht\strutbox]{figures/flags/running_svg-tex}}%
  \endgroup} {\begingroup\normalfont
  \raisebox{-.1\ht\strutbox}{\includegraphics[height=\ht\strutbox]{figures/flags/stable_svg-tex}}%
  \endgroup}
        {\begingroup\normalfont
  \raisebox{-.1\ht\strutbox}{\includegraphics[height=\ht\strutbox]{figures/flags/v2dir_svg-tex}}%
  \endgroup} {\begingroup\normalfont
  \raisebox{-.1\ht\strutbox}{\includegraphics[height=\ht\strutbox]{figures/flags/valid_svg-tex}}%
  \endgroup} \\

        495 & \textcolor{blue}{\textbf{Bananenbakker}} & \num{313283} & NL (100 km) & \as[link=true]{1102} & 37.86~MiB/s & 2018-05-04 & 5 days &
        {\begingroup\normalfont
  \raisebox{-.1\ht\strutbox}{\includegraphics[height=\ht\strutbox]{figures/flags/no-exit_svg-tex}}%
  \endgroup} {\begingroup\normalfont
  \raisebox{-.1\ht\strutbox}{\includegraphics[height=\ht\strutbox]{figures/flags/fast_svg-tex}}%
  \endgroup} {\begingroup\normalfont
  \raisebox{-.1\ht\strutbox}{\includegraphics[height=\ht\strutbox]{figures/flags/guard_svg-tex}}%
  \endgroup}
        {\begingroup\normalfont
  \raisebox{-.1\ht\strutbox}{\includegraphics[height=\ht\strutbox]{figures/flags/no-hsdir_svg-tex}}%
  \endgroup} {\begingroup\normalfont
  \raisebox{-.1\ht\strutbox}{\includegraphics[height=\ht\strutbox]{figures/flags/running_svg-tex}}%
  \endgroup} {\begingroup\normalfont
  \raisebox{-.1\ht\strutbox}{\includegraphics[height=\ht\strutbox]{figures/flags/stable_svg-tex}}%
  \endgroup}
        {\begingroup\normalfont
  \raisebox{-.1\ht\strutbox}{\includegraphics[height=\ht\strutbox]{figures/flags/v2dir_svg-tex}}%
  \endgroup} {\begingroup\normalfont
  \raisebox{-.1\ht\strutbox}{\includegraphics[height=\ht\strutbox]{figures/flags/valid_svg-tex}}%
  \endgroup} \\

        \bottomrule
    \end{tabular}

    \caption{Top-15 relays ranked by network adversary metric ($NM$)}
    \label{tab:net_top15}
\end{subtable}

    \begin{tabular}{llllllll}
        {\begingroup\normalfont
  \raisebox{-.1\ht\strutbox}{\includegraphics[height=\ht\strutbox]{figures/flags/exit_svg-tex}}%
  \endgroup} Exit & {\begingroup\normalfont
  \raisebox{-.1\ht\strutbox}{\includegraphics[height=\ht\strutbox]{figures/flags/fast_svg-tex}}%
  \endgroup} Fast & {\begingroup\normalfont
  \raisebox{-.1\ht\strutbox}{\includegraphics[height=\ht\strutbox]{figures/flags/guard_svg-tex}}%
  \endgroup}
        Guard &
        {\begingroup\normalfont
  \raisebox{-.1\ht\strutbox}{\includegraphics[height=\ht\strutbox]{figures/flags/hsdir_svg-tex}}%
  \endgroup} HSDir & {\begingroup\normalfont
  \raisebox{-.1\ht\strutbox}{\includegraphics[height=\ht\strutbox]{figures/flags/running_svg-tex}}%
  \endgroup} Running &
        {\begingroup\normalfont
  \raisebox{-.1\ht\strutbox}{\includegraphics[height=\ht\strutbox]{figures/flags/stable_svg-tex}}%
  \endgroup}
        Stable & {\begingroup\normalfont
  \raisebox{-.1\ht\strutbox}{\includegraphics[height=\ht\strutbox]{figures/flags/v2dir_svg-tex}}%
  \endgroup} V2Dir & {\begingroup\normalfont
  \raisebox{-.1\ht\strutbox}{\includegraphics[height=\ht\strutbox]{figures/flags/valid_svg-tex}}%
  \endgroup} Valid \\
    \end{tabular}

    \caption{Top-15 relays ranked by each metric for the exclusion of October 2022.
    \textcolor{blue}{\textbf{Blue and bold}} relay names appear in both \cref{tab:net_top15,tab:relay_top15}.
    The first column indicates the rank of the relay in the consensus when sorted by the metric value.
    GeoIP data from~\citet{maxmind_geolite}.}
\end{table*}

For each relay, \cref{tab:relay_top15,tab:net_top15} report the approximate geographic location (based on the IP address),
the \ac{as} hosting the machine (also based on the IP address), the highest bandwidth reported by the relay for any
10-second period over the past 5 days, the relay's age and uptime, and its flags.
Relays from the same family are indicated by an icon next to their names.
Blue, bold font indicates that the relay appears in both~\cref{tab:relay_top15,tab:net_top15}.
Although not all of these attributes are used to compute the metrics, they may help readers understand the relay's profile.

All information displayed in \cref{tab:relay_top15,tab:net_top15} is publicly available.
\section{\ac{as} adversaries ranking}
\label{appendix:as-adversaries-ranking}

\begin{figure*}
    \centering
    \begin{subfigure}[b]{0.24\textwidth}
        \centering
        \includegraphics[width=\textwidth]{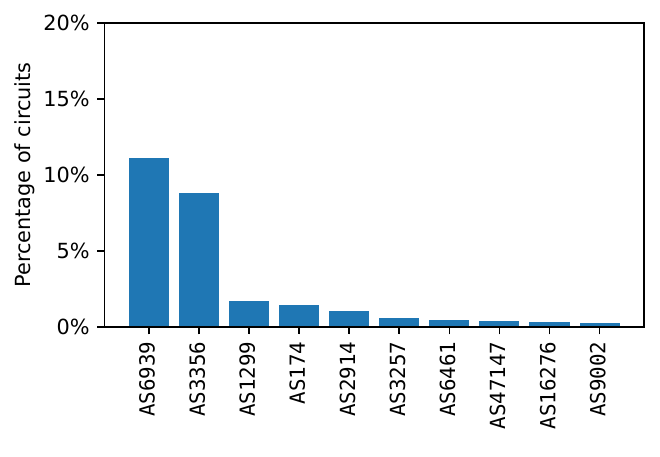}
        \caption{December 2021}
        \label{fig:sus-ases:2021-12}
    \end{subfigure}
    \begin{subfigure}[b]{0.24\textwidth}
        \centering
        \includegraphics[width=\textwidth]{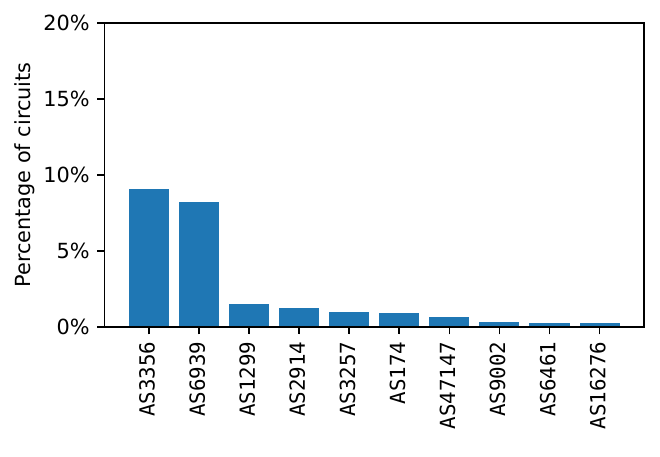}
        \caption{March 2022}
        \label{fig:sus-ases:2022-03}
    \end{subfigure}
    \begin{subfigure}[b]{0.24\textwidth}
        \centering
        \includegraphics[width=\textwidth]{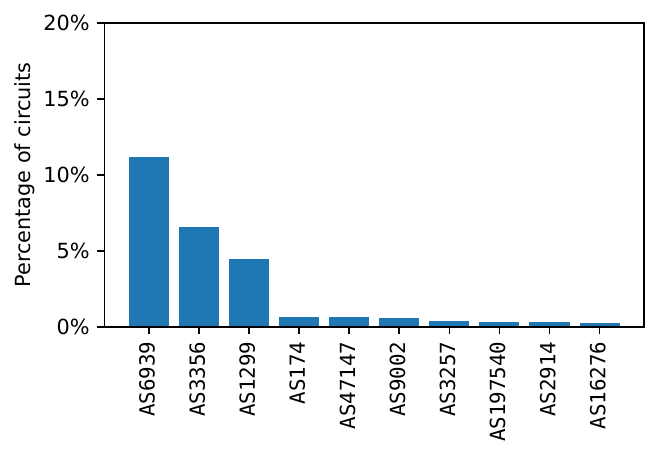}
        \caption{October 2022}
        \label{fig:sus-ases:2022-10}
    \end{subfigure}
    \begin{subfigure}[b]{0.24\textwidth}
        \centering
        \includegraphics[width=\textwidth]{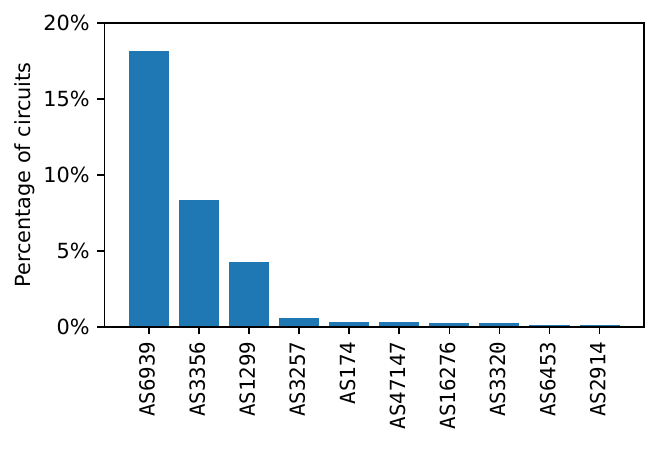}
        \caption{May 2023}
        \label{fig:sus-ases:2023-05}
    \end{subfigure}
    \caption{Top-10 \acp{as} capable of observing circuits on both ends for the four exclusion dates considered.
    The percentage of circuits is the fraction of simulated circuits on which the \ac{as} is found on both ends.}
    \label{fig:sus-ases}
\end{figure*}

\Cref{fig:sus-ases} shows which \acp{as} are in a position to observe both ends of a circuit.
Note that the top-2 (\as[link=true]{6939} and \as[link=true]{3359}) and the top-3 (\as[link=true]{6939},
\as[link=true]{3359}, and \as[link=true]{1299}) are stable across the four dates considered.
The \acp{as} identified by our simulations are consistent with the top-ranked infrastructure \acp{as} identified
by~\citet{ipinfo_asn}.

\end{document}